\documentclass[sigconf]{acmart}

\usepackage{subfig}

\usepackage{enumitem}

\usepackage{xcolor}

\AtBeginDocument{%
  \providecommand\BibTeX{{%
    \normalfont B\kern-0.5em{\scshape i\kern-0.25em b}\kern-0.8em\TeX}}}

\copyrightyear{2022}
\acmYear{2022}
\setcopyright{acmcopyright}\acmConference[DIS '22]{Designing Interactive Systems Conference}{June 13--17, 2022}{Virtual Event, Australia}
\acmBooktitle{Designing Interactive Systems Conference (DIS '22), June 13--17, 2022, Virtual Event, Australia}
\acmPrice{15.00}
\acmDOI{10.1145/3532106.3533526}
\acmISBN{978-1-4503-9358-4/22/06}

\begin{document}


\title[DramatVis Personae]{\textsc{DramatVis Personae}: Visual Text Analytics for Identifying Social Biases in Creative Writing}


\author{Md Naimul Hoque}
\email{nhoque@umd.edu}
\affiliation{
    \institution{University of Maryland, College Park}
    \city{College Park}
    \state{MD}
    \country{USA}
}

\author{Bhavya Ghai}
\email{bghai@cs.stonybrook.edu}
\affiliation{
    \institution{Stony Brook University}
    \city{Stony Brook}
    \state{NY}
    \country{USA}
}

\author{Niklas Elmqvist}
\email{elm@umd.edu}
\affiliation{
    \institution{University of Maryland, College Park}
    \city{College Park}
    \state{MD}
    \country{USA}
}

\renewcommand{\shortauthors}{Hoque, et al.}


\begin{abstract}
    Implicit biases and stereotypes are often pervasive in different forms of creative writing such as novels, screenplays, and children's books. To understand the kind of biases writers are concerned about and how they mitigate those in their writing, we conducted formative interviews with nine writers. The interviews suggested that despite a writer's best interest, tracking and managing implicit biases such as a lack of agency, supporting or submissive roles, or harmful language for characters representing marginalized groups is challenging as the story becomes longer and complicated. Based on the interviews, we developed \textsc{DramatVis Personae} (DVP), a visual analytics tool that allows writers to assign social identities to characters, and evaluate how characters and different intersectional social identities are represented in the story. To evaluate DVP, we first conducted think-aloud sessions with three writers and found that DVP is easy-to-use, naturally integrates into the writing process, and could potentially help writers in several critical bias identification tasks. We then conducted a follow-up user study with 11 writers and found that participants could answer questions related to bias detection more efficiently using DVP in comparison to a simple text editor. 
    
\end{abstract}



\begin{teaserfigure}
    \centering
    \includegraphics[width=0.9\textwidth]{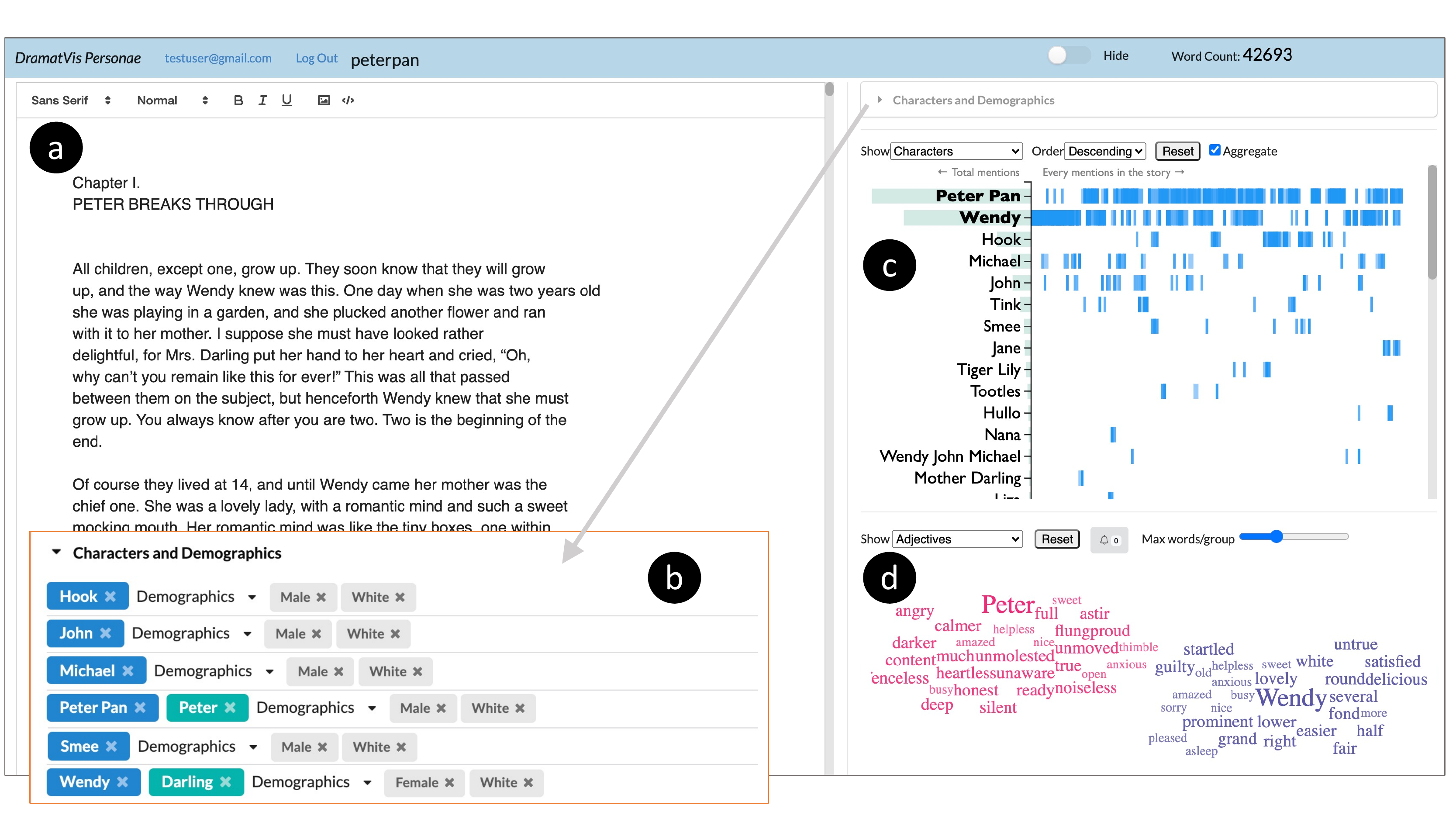}
    \caption{\textbf{Overview of \textsc{DramatVis Personae} (DVP)}.
    (a) Rich text editor.
    (b) Characters and Demographics panel for listing characters identified by the system, merging aliases, and assigning social identities to characters.
    (c) Timeline representation of the story (\textit{Peter Pan} by J.\ M.\ Barrie (1911)) showing every mentions of characters as well as the total number of mentions of characters.
    (d) Word zone~\cite{hearst2019evaluation} showing sample adjectives used for the selected characters (Peter Pan and Wendy).}
    \Description{A picture showing four different parts of DramatVis Personae (DVP) interface.}
    \label{fig:teaser}
\end{teaserfigure}

\maketitle

\section{Introduction}
Gandalf.
Elizabeth Bennet. 
Hermione Granger and Ron Weasley.
Atticus Finch.
Anomander Rake, Lord of Moon's Spawn and Son of Darkness.
Holly Golightly, Lisbeth Salander, and Hannibal Lecter: literature lives and dies by its characters.
Heroes and anti-heroes, villains and bad guys, innocent bystanders or willing accomplices---fiction is arguably about conjuring more or less complete humans out of whole cloth and then providing audiences with the emotional release of \textit{catharsis}, often by having these characters go through hell and high water.
Therein lies also the secret of great literature: creating believable, nuanced, and multidimensional characters that spring out of the written page and come to life in the reader's mind.
Achieving this is no mean feat, particularly when considering that truly great fiction often requires a diverse, inclusive, and just treatment of its cast of characters; its \textit{dramatis personae}.

Creative writing or storytelling can also be seen as a reflection of our societal beliefs, while at the same time societal beliefs
can be influenced by stories~\cite{correll2007influence}.
Thus, it is imperative for written stories to not promote biased representation of minority and marginalized groups.
However, current literature is filled with tired, unoriginal, and sometimes harmful stereotypes related to race, gender, sexuality, ethnicity, and age, such as the trope of the angry African-American woman, the studious Asian person, or the helpless damsel in distress~\cite{beckett2010away, layne2015zebra, hoyle-etal-2019-unsupervised, fast2016shirtless, gdblack2019, gdfilms2008, joseph2017girls, kolbe1981sex}.

Change is coming, with various organizations, institutes, writers, and the publishing community working continuously to raise awareness against biases in creative writing.
For example, the Geena Davis Institute regularly publishes reports of gender and racial representation in Hollywood and foreign creative materials~\cite{gdfilms2008, gdblack2019}.
Twitter hashtag ``OwnVoices'' promotes writers from marginalized groups who write about their community.
Professional writers now seek feedback from ``Sensibility Readers'', a group of readers who especially look for harmful stereotypes before the material is published.
On the computational front, Natural Language Processing (NLP) has been used to measure stereotypes in creative writing.
Many of these studies have helped us understand how stereotypes operate in culture by analyzing corpus containing millions of books, a scale much larger than any previous analysis~\cite{hoyle-etal-2019-unsupervised, norberg2016naughty, fast2016shirtless}.

As a result of these efforts, writers are increasingly becoming aware of harmful biases and stereotypes, and we see reports of better representation in recent creative materials~\cite{parity_children, parity_films, uclareport2020}.
However, little is known about writers' current practices for addressing biases in their writing. Given that biases often take nuanced, complicated, and intersectional forms that can be hard to detect, we speculate that computational support in this regard can help writers detect biases and write more inclusive and representative materials.
Motivated by that, this paper seeks to understand creative writers' current practices for tackling biases and how computational tools can potentially help them in this regard.
To inform our research, we conducted formative interviews with nine creative writers with published stories in their portfolios.
The interviews revealed that writers are mostly concerned about two types of biases: (1) \emph{Lack of agency} for minority characters (e.g., a female character introduced only to forward the plot for a male protagonist); (2) \emph{Stereotypes} encoded in how characters are \emph{described} and the \emph{actions} they take in the story (e.g., a female character described as beautiful and homely).
Writers mentioned that they actively look for such biases in their stories.
However, these biases are often implicit and unconscious and difficult to wheedle out even for the best and most self-reflective of authors.
The process is even more challenging for longer and complicated stories where many characters take intersectional identities. 

Based on the findings of the formative interviews, we designed \textsc{DramatVis Personae}{} (DVP), a web-based visual analytics system to help writers identify stereotypes in creative writing.
DVP is designed to integrate smoothly with the writer's own creative process, allowing them to analyze existing literature for research, upload their written content as it becomes available
or even write in the tool itself, and then having its text analytics and visualizations update in real time.
Using NLP methods such as entity recognition, co-reference resolution, and dependency parsing, DVP automatically detects characters in the text and collects data about them as the story progresses, including their aliases, mentions, and actions.
The author can then furnish demographic information for each character, such as their age, ethnicity, gender, etc.
The DVP dashboard uses this continually growing dataset to visualize the presence of characters and social identities over time. 

After our initial design and implementation, we approached writers from the formative interviews and conducted think-aloud sessions using the tool.
During a hands-on evaluation session conducted over videoconferencing, one writer was asked to use the tool to write a short story given a specific writing prompt.
Other participants used the tool to evaluate their own existing stories.
We observed their performance and then interviewed them with regard to their experience.
All participants expressed positive sentiment about the DVP tool, claiming that it helped them get a better grip of their characters and their story arcs throughout the process.
In particular, all participants appreciated that the tool managed and visualized character demographics, suggesting that the tool is helpful in writing a more nuanced and equitable story.
We further conducted a user study with $11$ participants to evaluate the effectiveness of DVP in detecting biases.
The study revealed that participants could answer questions related to bias detection more efficiently using DVP in comparison to a simple text editor. 

In sum, we claim the following contributions with this work: 
(1) findings on how to support the creative writing process by mitigating implicit bias, via an interactive interview session with nine fiction writers; 
(2) a visual analytics tool called \textsc{DramatVis Personae}{} (DVP) for supporting both online creative writing as well as offline analysis of fiction;
(3) results from three separate think-aloud sessions of deploying DVP with creative writers in both story generation as well as story analysis settings;
and (4) results from a summative user study, outlining the effectiveness of DVP in detecting biases and stereotypes. 

\section{Background and Related Work}

In this paper, we focus on ``creative writing,'' or the production of the written artifacts capturing the narrative, such as the book manuscript, fiction, or short story.
Creative writing falls under the umbrella of ``creative storytelling'' since authors are essentially telling stories through writing.
The rest of this section is designed to discuss research around creative writing, bias in creative writing, NLP, and text and literary visualization.

\subsection{Bias in Creative Writing}
\label{sec:story-bias}

Stereotypes in the form of art often mirror the problems, issues, thinking, and perception of different social groups in society~\cite{ross2019media, tsao2008gender}.
They can further reinforce biases and stereotypes against minority and marginalized groups in society~\cite{correll2007influence}.
The presence of biases and stereotypes, especially gender and racial bias, has been reported ubiquitously in different forms of creative writing.
We provide a brief overview of research in this area below. 

Many researchers have shown the prevalence of gender stereotypes in children's books, dating back to the early 1970s~\cite{flerx1976sex}.
Since then, several studies have reported that males are often portrayed as active and dominating, while females are instead described as passive and soft~\cite{kolbe1981sex, narahara1998gender, paynter2011gender}.
Other studies have found the presence of racial bias~\cite{layne2015zebra}, stereotypes against disability~\cite{beckett2010away}, and occupation~\cite{hamilton2006gender} in children's books.
Researchers have argued that the presence of such stereotypes in children's books is severely problematic as children are susceptible to inheriting stereotypes at an early age~\cite{kolbe1981sex, tsao2008gender}.
While the situation is improving (i.e., females are portrayed with more active roles in recent children's books) due to increased social awareness, the improvement is not significant~\cite{peterson1990gender}, and there are still reports of the prevalence of different stereotypes in children's books~\cite{adukia2021we, hamilton2006gender, CLPE}. 

Another form of creative writing medium that has been heavily criticized for promoting stereotypes is movie scripts.
The \textit{Geena Davis Institute} regularly publishes reports of gender and racial representation in Hollywood movies and is a valuable resource for current representational problems.
The institute has found underrepresentation and misrepresentation of females~\cite{gdfilms2008} and Black or African American females in Hollywood~\cite{gdblack2019}.
Beyond Hollywood, researchers have found similar sorts of biases in television shows and movies in other countries.
Emons et al.~\cite{emons2010he} found stereotypes in gender roles of males and females in U.S.-produced Dutch TV shows, misrepresenting females in Dutch society.
Madaan et al.~\cite{madaan2018analyze} has shown the existence of gender biases in Bollywood movie scripts. Similar to movie scripts, many studies have shown how biases and stereotypes operate implicitly in news articles and how they adversely affect the audience~\cite{correll2007influence, entman1992blacks, ramasubramanian2007activating, valentino1999crime}.

Finally, newer forms of writing such as blogs, online writeups, and social media posts are rife with harmful stereotypes.
Fast et al.~\cite{fast2016shirtless} analyzed fiction written by novice writers in the online community Wattpad and found it to be rampant with common gender stereotypes.
Joseph et al.~\cite{joseph2017girls} analyzed forty-five thousand Twitter users who actively tweeted about the Michael Brown and Eric Garner tragedies.
Their method can quantify semantic relations between social identities.
Other work discussed the impact of stereotypes in Reddit~\cite{ferrer2020discovering}, Facebook~\cite{matamoros2017platformed}, and U.S.\ history books~\cite{lucy2020content}.

All the above-mentioned research has been instrumental in raising awareness among writers, directors, and the general audience, a critical step towards equality.
As a result, there are reports of better representation and inclusivity in recent years~\cite{parity_children, parity_films, uclareport2020}.
Originating on Twitter in 2015, ``OwnVoices'' has become  a  campaign for promoting writers from diverse backgrounds writing about their experiences and cultures.
We strongly believe that empowering writers from historically excluded groups is tremendously important for our society.
At the same time, we believe writers from all backgrounds need to be cautious when writing creative materials that represent a culture or social identity and invest efforts to learn about relevant communities for correct portrayals.
In fact, we believe our tool will be most useful for writers who want to ensure that they represent other communities correctly and do not propagate stereotypes.
We intentionally avoided identifying a specific user group for our tool as we acknowledge that identities are intersectional, meaning a person who is considered privileged in one dimension of their identity, may well be a minority in another dimension. 
Rather, we strove to include writers from diverse backgrounds in the design of this tool.
We believe our tool can assist in the production of more inclusive and representative writings that will have a positive impact on society.

\subsection{Computational Support for Creative Writing}

\label{background:process}


Beyond the everyday-use text processors such as Microsoft Word, there are several professional and open-source software available to writers for helping them in guiding character development.
Scrivener~\cite{scrivener} is a paid service for organizing stories.
It allows flexible page breaking, adding synopsis and notes to each section, and easy merging or swapping between sections.
It also has a distraction-free writing mode where everything else on the computer is tuned out.
Granthika~\cite{granthika} is a similar sort of paid service that helps writers in tracking characters and events in a story.
It lets users integrate knowledge into the system as they write, and then use that knowledge for tracking in a timeline.
It also allows writers to apply causal constraints to the events and people in a story (e.g, ``The inquest must happen after the murder''). Grammarly is used for checking grammatical errors.

Over the years, NLP research has developed and refined different techniques such as coreference resolution, named entity recognition (NER), dependency parsing, sentiment analysis, etc, which can be instrumental in analyzing stories.
For example, NER helps extract named entities such as person names, organizations, locations, etc from the text.
In our context, this can help identify different characters of a story~\cite{dekker2019evaluating}.
Dependency parsing finds relationships between words.
This can help identify different adjectives/verbs linked to a character in a story~\cite{mitri2020story}. 
Similarly, coreference resolution can help track the representation of different characters and their demographic groups across the storyline~\cite{kraicer2019social}.     
Other studies in the NLP literature have specifically focused on analyzing stories.
This includes segmenting stories by predicting chapter boundaries~\cite{pethe2020chapter}, recognizing flow of time in a story~\cite{kim2020time}, analyzing emotional arc of a story~\cite{reagan2016emotional}, extracting character networks from novels~\cite{labatut2019extraction}, etc.
Using such methods, researchers have conducted large-scale analysis of books, and stories~\cite{norberg2016naughty, pearce2008investigating, hoyle-etal-2019-unsupervised, labatut2019extraction, fast2016shirtless, joseph2017girls}.

Finally, the invent of powerful generative language models such as GPT-3~\cite{brown2020language} have fueled research on human-AI collaboration for creative writing.
Many tools in this area generate stories iteratively: a writer provides an initial prompt for the story; AI generates a section of the story automatically; the writer edits the generated story and provides further prompts and so on~\cite{clark2018creative, coenen2021wordcraft, lee2022coauthor}.
Recently, Chung et al.~\cite{talebrush} proposed TaleBrush, an interactive tool where AI can generate a narrative arc based on a sketch drawn by a writer that outlines the expected changes in the fortunes of characters.
Other works include IntroAssist~\cite{hui2018introassist}, a web-based tool that helps entrepreneurs to write introductory help-requests to potential clients, investors, and stakeholders; INJECT~\cite{maiden2018making}, a tool for supporting journalists explore new creative angles for their stories under development; and an interactive tool proposed by Sterman et al.~\cite{sterman2020interacting} that allows writers to interact with literary style of an article. 



While these works provide necessary background for the technical design of our tool, none of these works have features to add social identities to characters and investigate potential biases.
Additionally, prior works have primarily used textual descriptions for summarizing and communication.
While that is helpful, we utilize data visualization, a visual communication medium, for gathering insights from the information extracted by the tool efficiently.
.



\subsection{Visualization for Text and Literature}

Harking back to some of the original approaches to visualizing ``non-visual'' text documents~\cite{DBLP:conf/infovis/WiseTPLPSC95}, data visualization has long been proposed as an alternative to reading through large document corpora~\cite{DBLP:journals/widm/AlencarOP12, Qihong2014, DBLP:conf/kes/SilicB10}. 
For example, the investigative analytics tool Jigsaw~\cite{DBLP:conf/ieeevast/StaskoGLS07} is often styled as a ``visual index'' into a document collection; while it is not a replacement for reading, it provides a linked collection of entities and documents for easy overview and navigation.
This is generally also true for document and text visualization as a whole; the goal is to be able to ``see beyond'' the raw text into content, structure, and semantics~\cite{DBLP:journals/widm/AlencarOP12}.

Beyond simplistic text visualization techniques such as word clouds, data visualization can become particularly powerful when applied to entire documents~\cite{Qihong2014}.
These ideas can also be used for literary analysis of fiction and poetry~\cite{DBLP:journals/cgf/CorrellWG11}.
For example, Rohrer et al.~\cite{DBLP:conf/infovis/RohrerSE98} used implicit 3D surfaces to show similarities between documents, such as the work of William Shakespeare.
Similarly, Keim and Oelke propose a visual fingerprinting method for performing comparative literature analysis~\cite{DBLP:conf/ieeevast/KeimO07}.
McCurdy~\cite{DBLP:journals/tvcg/McCurdyLCM16} present a organic linked visualization approach to scaffolding close reading of poetry.
The literary tool Myopia~\cite{DBLP:conf/dihu/ChaturvediGMAH12} uses color-coded entities to show the literary attributes of a poem for readers.
Abdul-Rahman et al.~\cite{DBLP:journals/cgf/Abdul-RahmanLCMMWJTC13} also apply data visualization to poetry.
Our work here is inspired by, if not the design and implementation, then at least the motivation of these tools; however, in comparison, our goal is to support the creative processing while focusing on identifying and mitigating implicit social bias.

Finally, visualization can also be applied to the stories themselves rather than the actual text.
XKCD \#657,\footnote{\url{https://xkcd.com/657/}} titled \textit{Movie Narrative Charts}, shows temporal representations of plotlines in five movies, including the original \textit{Star Wars} trilogy (1977--1983), \textit{Jurassic Park} (1993), and the complete \textit{Lord of the Rings} movie trilogy (2001--2003).
Liu et al.~\cite{DBLP:journals/tvcg/LiuWWLL13} propose an automated approach to generating such storyline visualizations called StoryFlow.
Tanahashi and Ma discuss design considerations for best utilizing storylines~\cite{DBLP:journals/tvcg/TanahashiM12}.
Some effort has also been directed towards minimizing crossings in storyline visualizations~\cite{DBLP:conf/gd/GronemannJLM16}.
VizStory~\cite{DBLP:conf/taai/HuangLS13} takes a very literal approach to visualize stories by identifying segments and themes and then searching the web for appropriate representative images.
TextFlow~\cite{Cui2011} and ThemeDelta~\cite{DBLP:journals/tvcg/GadJGEEHR15} and related topic modeling visualizations can be used to automatically extract and visualize evolving themes in a document (or document collection) over time. 
StoryPrint~\cite{DBLP:conf/iui/WatsonSSGMK19} shows polar representations of movie scripts and screenplays in a fashion similar to Keim and Oelke's fingerprints; the authors note that this approach could also be used to support the creative writing process. StoryCurves~\cite{DBLP:journals/tvcg/KimBISGP18} shows non-linear narratives in movies on a timeline representation. Finally, Story Analyzer~\cite{mitri2020story} shows several visualizations representing summary statistics of a story.
Common for all of these storylines and plot visualization tools is that they are designed mostly for retrospective analysis and not for online creative writing.
None of these tools support bias identification in an interactive environment.
Nevertheless, we draw on all of these tools in our design of DVP.

\section{Formative Study}


To understand creative writers' current practices and challenges for addressing biases, we conducted semi-structured interviews with 9 creative writers.
The study was approved by our university's Institutional Review Board (IRB).

\begin{table*}[t!]
    \centering
    \caption{Participant demographics.}
    \Description{A table showing demographic information of nine writers.}
    \label{tab:participant}
    \begin{tabular}{llp{3.4cm}cp{8.5cm}c}
    \toprule
     \textbf{Id} & \textbf{Gender} & \textbf{Race} & \textbf{Age} & \textbf{Expertise} & \textbf{Yrs Exp} \\ 
    \midrule
    W1 & Male & Asian & 30 & Short stories, poems, and blogs & 10  \\
    W2 & Male & Asian & 27 & Short stories, poems, and critiques & 12 \\
    W3 & Female & White & 26 & Novels (fiction/non-fiction), short stories, screenplays, poems, blogs, critiques, and fanfiction & 15 \\
    W4 & Female & White & 49 & Novels (fiction/non-fiction), and roleplaying games & 25 \\
    W5 & Female & Black or African-American & 44 & Picture books and books for beginning readers & 6 \\
    W6 & Non-binary & Prefer not to respond & 36 & Novels (fiction/non-fiction), short stories, and poems & 10\\
    W7 & Female & Asian & 25 & Short stories and poems & 15\\
    W8 & Male & Asian & 25 & Screenplays, blogs, and critiques & 12\\
    W9 & Female & White & 34 & Screenplays and poems & 15\\ \bottomrule
    \end{tabular}
\end{table*}

\subsection{Participants}

We recruited 9 creative writers through advertisements to social media such as Twitter and Facebook, local mailing lists, and university mailing lists.
Our inclusion criteria included prior experiences in creative writing such as novels, short stories, screenplays, etc, and familiarity with writing in text editors. All participants had published materials in their portfolios.
Table~\ref{tab:participant} presents participants' demographic information.
Each participant received an Amazon gift card worth \$15 for their participation.


\subsection{Procedure}

We conducted the interviews over Zoom.
Each interview lasted around 1 hour and was divided into three parts.
First, after gathering informed consent, we asked the writers to share their perspectives on bias in creative writing. In the second part, we asked the writers about the challenges they faced in addressing biases in their writing, and their current approach for overcoming biases. Finally, writers brainstormed with the study administrator for outlining the potentials and requirements for digital tools that might help them in managing biases and stereotypes. We provide the questionnaire for the interviews as a supplement.

\subsection{Analysis}

We created anonymized transcript for each interview from the recorded audio.
Two authors of this paper open-coded the transcripts independently.
A code was generated by summarizing relevant phrases or sentences from the transcripts with a short descriptive text.
Both coders then conducted a thematic analysis~\cite{braun2006using} to group related codes into themes.
Throughout this process, the coders refined the themes and codes over multiple meetings by discussing disagreements and adjusting boundaries, scopes, and descriptions of the codes and themes.
The open codes and themes were also regularly discussed with the full research group.
We present the findings from the interviews next.





\subsection{Findings}

Our findings relate to the topics of implicit bias in creative writing, mitigating bias, and the potential for computational support for such issues.

\subsubsection{Current State of Biases in Creative Writing}
The publishing and writing community have progressed towards  an inclusive and representative environment; however bias in creative writing is still a present and complicated problem (W1-W9). W3 mentioned that publishers now encourage writers to seek feedback from ``Sensitivity Readers'' for potential misrepresentation.
W4 lauded the recent Twitter hashtag ``\#ownvoices'' that started the discussion about the importance of writers from marginalized groups writing about their community. Despite these efforts, there is still a need for diverse writers while writers from all communities need to be conscious about how their writing represents different social groups (W1-W9).
W5 said: \textit{``One recent study\footnote{https://www.theguardian.com/books/2020/nov/11/childrens-books-eight-times-as-likely-to-feature-animal-main-characters-than-bame-people} found that there were more characters that feature animals than there were of all of the minorities.
    I mean, it is horrendous, to be honest with you, in 2021... and it is really sad.''} W7 emphasized the importance of research (e.g., reading literature, understanding culture) before writing about characters that represent marginalized groups.

\subsubsection{Bias 1: Lack of agency for minority characters}
\label{bias1}
A critical challenge for writers is to ensure that characters representing minority groups have impact in the story and are not sidelined. W3 said: \textit{``I believe it is Gail Simone.
    She is a comic book writer and she started this website called ``Women in Refrigerators'' because there is a very famous storyline from a Green Lantern comic (vol. 3, \#54) where the Green Lantern comes home to find that his girlfriend has been murdered and shoved in his refrigerator.
    And the only reason that the woman existed was to be killed.
    So it has become a name for this type of killing of female characters to advance the man's plot, which is unfortunately very common.''} W3 also referred to the \textit{sexy lamp test}, which tests if a female character of a story is replaced with a sexy lamp whether the story still makes sense.
W4 mentioned the \textit{Bechdel} test, which asks whether a story features at least two women who talk to each other about something other than a man. 



To address this type of bias, writers employ several self-evaluation techniques. They mentally track the presence of characters as well as their social identities in the story~(\textbf{N1|W3-7, W9}).
This frequently requires them to go back and forth between different parts of the story and read through them (\textbf{N2|W3-7, W9}). While reading, writers investigate interactions between characters (\textbf{N3|W3-7}), and how actions of the characters reflect on the social identities they represent (\textbf{N4|W3-5, W7}).  However, this process becomes challenging and tiresome as the stories become longer and complicated (W3-7, W9). Moreover, the identities often take intersectional form (e.g., a Female African-American character) which makes it even more mentally demanding to track them. It is worth noting here that writers also apply these methods to analyze existing literature as part of their research; although often that process is not as thorough as correcting their own work and depends mostly on high-level subjective understanding (W2-4, W7-9). Such research helps them understand representation of different social identities as portrayed in existing literature (W2-4, W7-9).

\subsubsection{Bias 2: Stereotypes in describing characters}
\label{bias2}
Another form of bias is how characters are described in the story. For example, W1 mentioned that female characters are often described as homely, beautiful, and sacrificial.  W7 said: \textit{``One example would be Truman Capote's Breakfast at Tiffany's. I read the book as a teenager, and it's very beautiful and well-written overall. But one example of a stereotyped character is Holly's neighbor Mr.\ Yunioshi. He is a side character but is portrayed simply as the irritable neighbor who always tells Holly off for forgetting her keys. This depiction of East Asian characters as grumpy old people is very stereotypical and common in Western literature.''}

To address this type of bias, writers remain careful while writing and examine the actions (\textbf{N4 from above| W1-4, W9}) and descriptors (\textbf{N5| W1-3, W7-9})  they use for the characters and social identities. This again requires them to go back and forth between different parts to critically read the story (\textbf{N2 from above| W1-4, W7-9}), and constantly check the actions and descriptors. Writers employ similar evaluation techniques for analyzing existing literature (W2-4, W9).

\subsubsection{Potential and recommendations for an interactive tool}
\label{sec:tool_potential}

All participants were enthusiastic for an analytic tool for identifying potential stereotypes.
Critique groups and sensitivity readers provide important feedback to writers; however, they are usually available at the advanced stage of a formal publication. During writing, a tool supporting their self-reflective process would be helpful. The tool will also be helpful to a writer to analyze existing literature as a part of their research (W2-4, W7-9).

Writers also provided a few recommendation for such a tool. First, writers suggested that the tool could support story writing since the various bias mitigation strategies discussed in Section~\ref{bias1} and~\ref{bias2} are closely integrated into the writing cycle (\textbf{N6| W2, W4, W8}). 
Second, the tool could have a ``distraction-free'' mode, similar to Scrivener, where writers could concentrate only on writing if they want to  (W2-5, W8). Finally, writers suggested that the tool should \textbf{avoid suggesting interpretations of its own}; rather the analytic tool should support writers in \textbf{exploring their work} so that they can make informed decisions. Writers provided several reasons behind this suggestion. W1 said:\textit{``Language around different social groups is fluid and continuously evolving. What is acceptable today, may not be acceptable a few months from now.''} W6 said: \textit{``Creative writing can be subjective to a writer's own experience, or imagination, and may elicit intended stereotypes for the plot. I want to write about discrimination against my community which if flagged by a tool would be disappointing''}

\section{Design Guidelines}
\label{sec:design}

Here we outline and discuss the design guidelines and decisions that originated from the literature and formative study.

\paragraph{DG1. Nature of the Support: Exploratory.}
The first design decision we made is the nature of the support.
The formative study suggested several challenges that writers face in identifying biases and stereotypes.
They apply several evaluation techniques (Table~\ref{tab:needs}) that can be mentally demanding.
All participants were enthusiastic about a tool that will support them in this process.
However, based on the findings from Section~\ref{sec:tool_potential}, it was clear that the tool should avoid flagging any writing as inappropriate and suggesting how to reduce biases.
Suggesting interventions for reducing biases without a proper understanding of a writer's goal could seriously lower user trust in the system and be counter-productive.
Another potential concern is writers using such suggestions to write about cultures and identities outside their own identity without really investing efforts to learn about them. 
Considering all these, we decided that the primary goal of the tool would be to \textbf{help writers perform \textbf{N1} to \textbf{N6} easily, without any explicit recommendation}.
We decided to use interactive visualization for this purpose which is an effective way to summarize any form of abstract data and enable exploration in an interactive environment.

\begin{table}
    \centering
    \caption{Design needs identified from the formative study, based on the evaluation techniques performed by writers to identify biases.}
    \Description{A table presenting six design needs identified from the formative study.}
    \begin{tabular}{p{3.8cm} p{1.6cm} c}
       \toprule
       \textbf{Design Need}  &  \textbf{Purpose} & \textbf{Participant} \\
       \midrule
       \textbf{N1.} Evaluate presence of characters and social identities.   & Bias 1  & W3-7, W9 \\ \hline
       \textbf{N2.} Move between different parts of the story for reading.  & General  & W1-9 \\ \hline
       \textbf{N3.} Evaluate interactions between characters-characters and social identities-social identities.  & Bias 1  &  W3-7\\ \hline
       \textbf{N4.} Evaluate and compare actions of characters and social identities.  & Bias 1, Bias 2  & W1-7, W9 \\ \hline
       \textbf{N5.} Evaluate and compare descriptions of characters and social identities.  & Bias 2  &  W1-3, W7-9\\ \hline
       \textbf{N6.} Support writing in the tool.  & General  & W2, W4, W8 \\
       \bottomrule
    \end{tabular}
    \label{tab:needs}
\end{table}

\paragraph{DG2: Help writers evaluate agency for characters and social identities.}
The formative interviews revealed that writers are concerned about the lack of agency for characters that represent minority groups.
To ensure agency for characters, writers practice three evaluation techniques (\textbf{N1, N3, N4} from Table~\ref{tab:needs}).
Prior research has also used presence, and interactions between characters to quantify character agency~\cite{hoque2020toward, parity_films}.
Thus, our tool should support writers in these tasks.


\paragraph{DG3: Help writers evaluate stereotypes in describing characters.}
Another important concern raised during the formative interviews was the stereotypes used for characters. 
Writers mentioned that they search for possible stereotypes in the action characters take in the story (verbs, \textbf{N4}), and the words that describe the characters (adjectives, \textbf{N5}).
Previous research has also used verbs and adjectives to quantify biases and stereotypes~\cite{hoyle-etal-2019-unsupervised,  emons2010he, fast2016shirtless, 100years}.
Thus, our tool should help writers in these tasks.



\paragraph{DG4: Support writing and critical reading.}
\textbf{N2} and \textbf{N6} are not directly related to bias identification.
However, the formative study suggests support for these features is necessary to facilitate bias identification.
They will also enable bias identification in different stages of writing a story. For example, during the pre-writing (e.g., analyzing existing literature) or post-writing stage, a writer can use the support for reading (\textbf{N2}) and other bias identification methods for critical reading. During the writing stage, writers will require support for both writing (\textbf{N6}) and reading (\textbf{N2)}.

We also decided to include a ``distraction-free'' mode similar to Scrivener~\cite{scrivener} that will allow writers to write or read without any distraction and use our bias identification support only when they want to.
This will ensure that our tool does not obstruct the creative process for writers.

\paragraph{DG5: Inclusive design.}
Prior technological efforts in bias identification and mitigation often focused on a specific type of bias, or a binary view on identity (e.g., male, female as gender category)~\cite{100years, hoque2020toward, hoyle-etal-2019-unsupervised}.
However, the formative interviews suggest that writers may use diverse and intersectional identities for characters that may not conform to any pre-defined categories.
We anticipated that the tool would be demoralizing if a social identity that an author is writing about is not supported in the tool; especially since authors may themselves share that identity.
Thus, our design should be inclusive, and support social identities of any kind.


\paragraph{DG6: Easy-to-understand visualization and scalable computation}
To ensure accessibility for writers who are not familiar with complex data visualization paradigms, our tool should use easy-to-understand visualizations.
Finally, the formative study suggests the tool would be most useful as the story becomes larger and complicated.
Thus, our tool should support stories of large size and provide feedback to the writers in a short response time.
Similarly, the visual components should be able to show information extracted from large textual data.



%
















\section{\textsc{DramatVis Personae}{} (DVP)}


We present the visual components and analysis pipeline for the DVP tool below.

\subsection{Visual Interface}

In this section, we demonstrate four visual components (Figure~\ref{fig:teaser}) of the interface. We link the design guidelines from Section~\ref{sec:design} wherever applicable.
We also discuss design rationales and design alternatives considered during the development of the tool. The interface supports analysis for three different types of entity: (a) Characters, (b) Social Identities (e.g., Male, Female), and (c) Intersectional Social Identities (e.g., Muslim Males).
We use the term ``entity'' or ``entities'' to refer to the three entities together whenever applicable for brevity.
For demonstration purpose, we use several well-known western stories.

\paragraph{Text Editor}

The central component of DVP is a text editor (\textbf{DG4}). 
The text editor is equipped with traditional formatting features such as selecting fonts, font sizes, font-weights, etc.
We use QuillJS~\cite{QuillJS} as a rich text editor which has been widely used on the web including social media apps such as Slack, and Linkedin.


The visualizations are contained in a sidebar; a user can hide or show the sidebar by clicking the \textit{``Hide''} toggle button (\textbf{DG4}). This ensures that writers can see the visualizations whenever they want, but also can concentrate on writing and reading by hiding the sidebar whenever they want.

\begin{figure*}
    \centering
    \includegraphics[width = 0.95\textwidth]{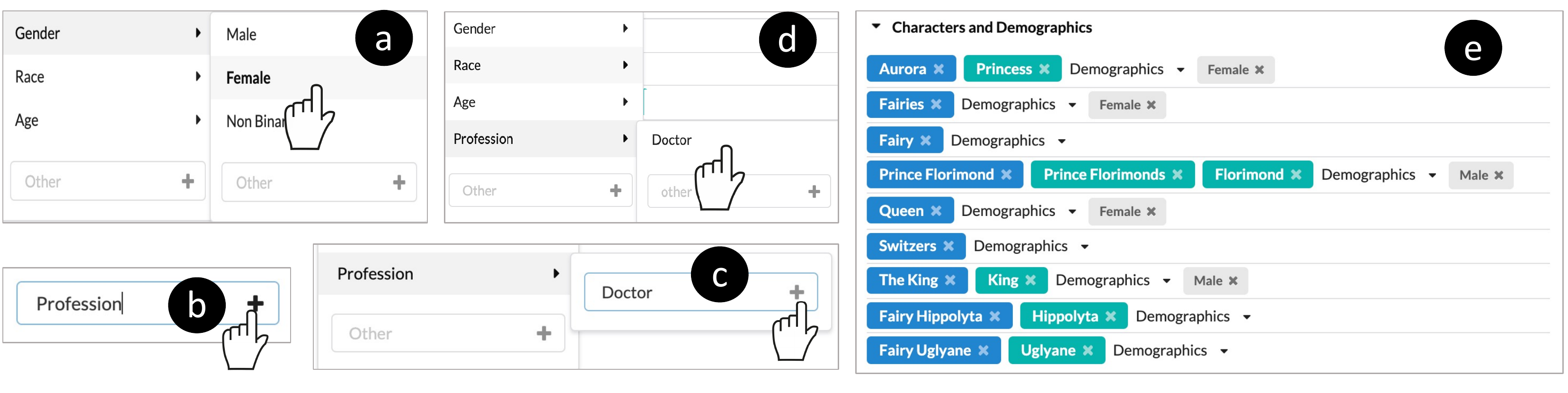}
    \Description{A picture with five different parts. The first four parts show how a user can dynamically add a social identity in the character and demographics panel. The last part shows an example demographics panel for the children's book, Sleeping Beauty. }
    \caption{\textbf{Characters and Demographics panel.} 
    (a) A dropdown menu with different social identities is available for each character in the Characters and Demographics Panel.
    (b) The dropdown can be extended dynamically.
    User can add a new identity (e.g., Profession).
    (c) User can then add categories under the newly added identity (e.g., Doctor).
    (d) The new Demographics dropdown with Doctor as a new profession.
    (e) An example demographics panel for the children's book, \textit{Sleeping Beauty}, retold by Arthur Quiller-Couch and Charles Perrault, freely available under Project Gutenberg~\cite{sleeping_beauty}.}
    \label{fig:demographics}
\end{figure*}

\paragraph{Characters and Demographics Panel}

The Characters and Demographics Panel lists all the named entities identified by our NLP pipeline. The panel supports several validation functionalities. First, a user can delete a character from the list in the case they do not wish to track that character, or if the character was wrongly identified by the NLP pipeline (e.g., an institute identified as a person). Second, a user can merge different names of the same character together (e.g., merging Peter Parker, and Spiderman as one character which might be identified as separate characters by the NLP pipeline). A user can use drag and drop for merging characters. Note that these validation functions are not necessarily a part of a writer's work process, but are needed to use the tool reliably given the nature of current NLP toolkits.


Each character in this panel has a dropdown named \textit{Demographics}. 
Using this dropdown, a user can add multiple social identities to each character (Figure~\ref{fig:demographics}a).
We populate the dropdowns with commonly used identities.
However, to support \textbf{DG5}, we made the dropdowns dynamically extendable.
A user can add any number of new identities in these dropdowns.
For example, Figure~\ref{fig:demographics}b and \ref{fig:demographics}c show how a user can add Profession as a new identity and Doctor as a profession in the dropdowns.

\paragraph{Timeline}

To support \textbf{DG2}, we designed a timeline representation of the story (see Fig.\ref{fig:teaser} (C)). The timeline is divided into two parts.
On the left side of the y-axis, we encode the total number of mentions for entities using bars encompassing the axis labels.
The y-axis can be sorted either in descending or ascending order.
A user can choose the sort order from the \textit{Order} dropdown.
The dropdown defaults to descending order for showing the prominent entities at the top.
On the right side of the y-axis, we show individual mentions for each character.
The x-axis represents a linear scale with a range $(1, S)$ where $S$ is the total number of sentences.
For a mention of an entity in sentence $s$, we draw a tile (rectangle) with width $(pos(s) + 0.5) - (pos(s) - 0.5) $ where $pos(s)$ represents the position of $s$ in the x-axis.
We used linear scale instead of ordinal scale to make the adjacent tiles connected and smoother.

\begin{figure*}
    \centering
    \includegraphics[width = 0.95\textwidth]{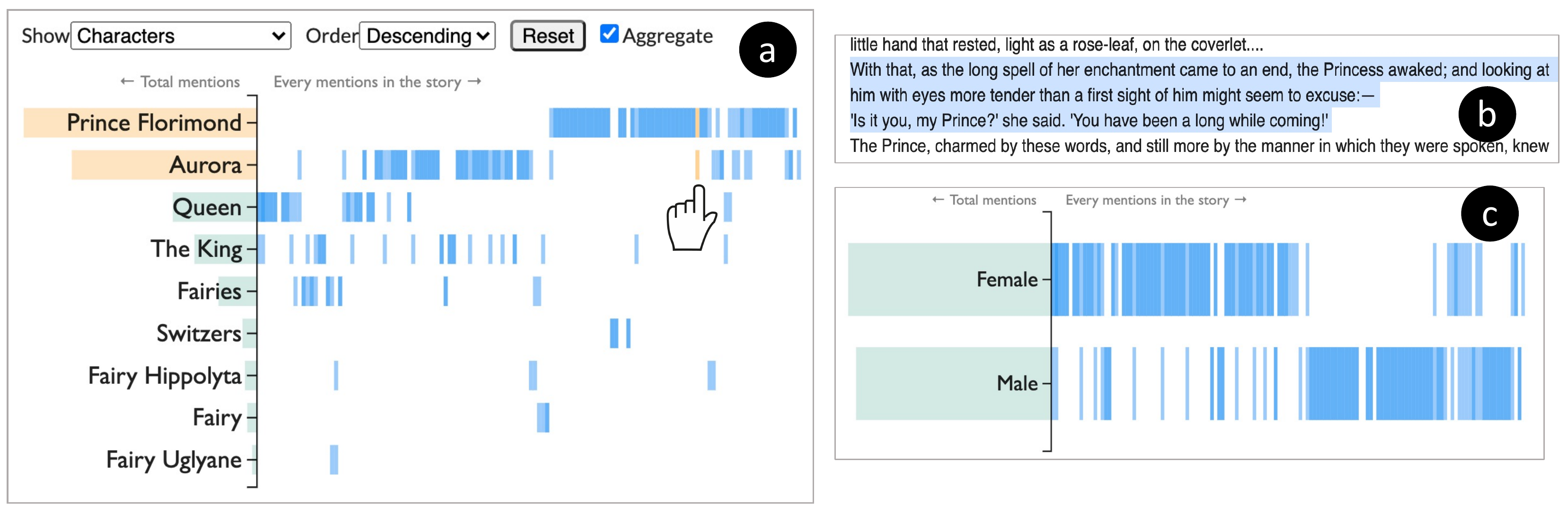}
    \Description{A picture with three subfigures showing different use cases for timeline visualization. }
    \caption{\textbf{Timeline interface.}
    (a) An example timeline based on the characters of \textit{Sleeping Beauty}.
    Note that even though the story is about Aurora, the sleeping beauty, it was Prince Florimond who had the most number of mentions.
    Note also the lack of overlaps between the timelines of two characters.
   (b) Upon hovering over a tile from the timeline, a user can see the relevant passage in the editor.
   The relevant passage in this case is the first scene after the Princess wakes up.
   (c) A user can also visualize the presence of different social identities in the timeline.
   We only used Male and Female for this example.
   Note that when aggregated, the female presence in the story is slightly larger than the male presence.}
   \label{fig:timeline}
\end{figure*}

For larger documents, due to space constraints, the tiles can become extremely small, making it difficult to interact with them.
To scale the visualization to larger documents (\textbf{DG6}), we added an \textit{Aggregate} option in the toolbar of the timeline (Figure~\ref{fig:timeline}a).
In the case of a document with more than 500 sentences $(S > 500)$, the option is automatically triggered, and the document is binned together to restrict the x-axis to range $(1,500)$.
At that point, the x-axis represents passages, instead of sentences.
The user can disable the aggregate option and see the timeline in terms of a single sentence.
As an alternative, we considered making the x-axis scrollable.
The tile size would have remained the same for any documents in that design.
However, we noticed that this design does not provide a full overview of the timeline together and it becomes difficult to compare different parts of the timeline as the user scrolls left and right in the timeline.

Figure~\ref{fig:timeline}a shows an aggregated version of the timeline for the children's book, \textit{Sleeping Beauty}, retold by Arthur Quiller-Couch and Charles Perrault, freely available under Project Gutenberg~\cite{sleeping_beauty}. The opacity of a tile represents the number of times an entity was mentioned in a particular passage. A user can hover over any tile and see the relevant passage, highlighted in the text editor (Figure~\ref{fig:timeline}b). This facilitates easily going back and forth between different parts of the story (\textbf{DG4}).

\begin{figure*}
    \centering
    \includegraphics[width = 0.9\textwidth]{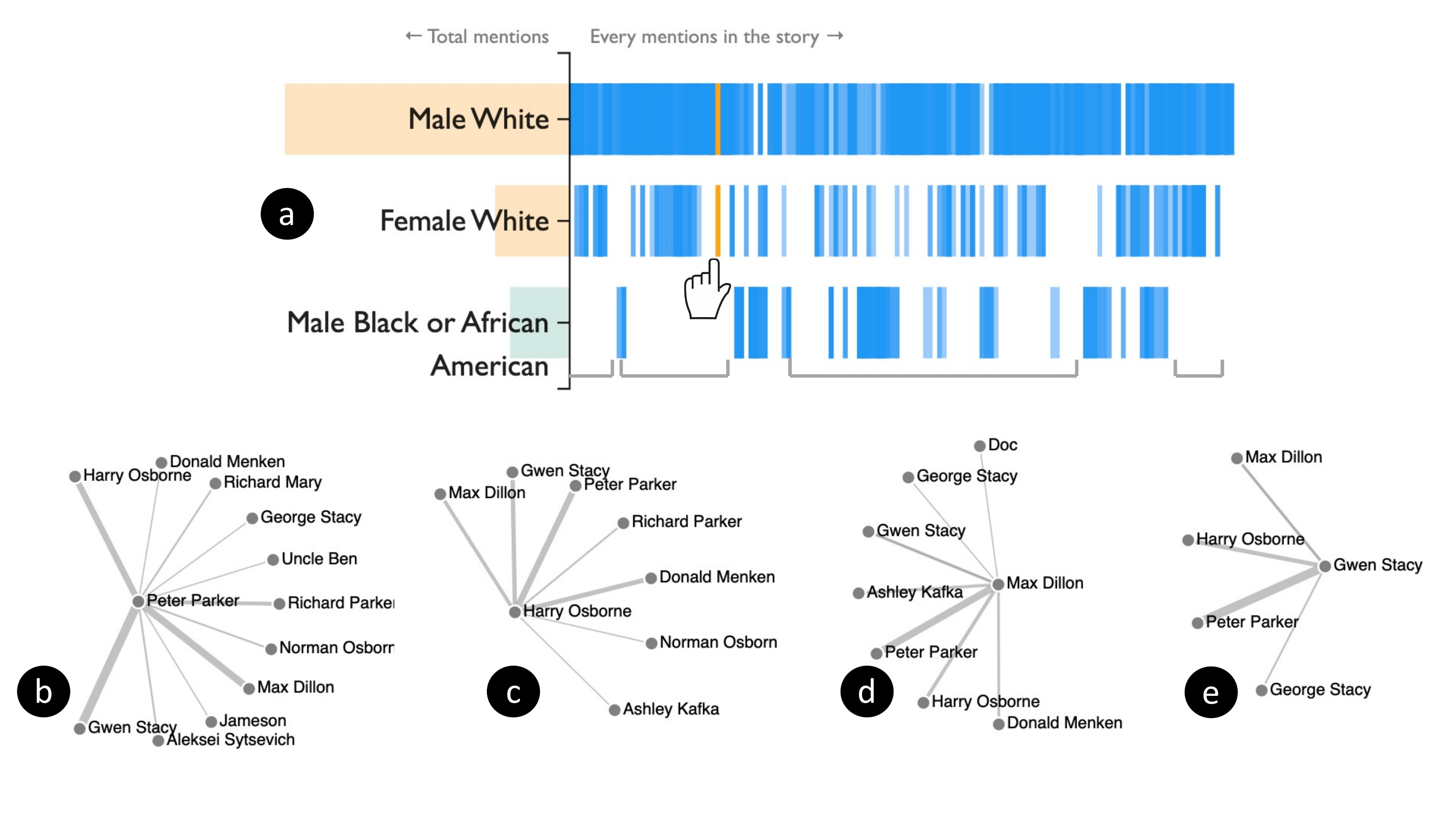}
    \Description{A picture showing a timeline and character interaction graphs for the movie The Amazing Spiderman 2.}
    \caption{\textbf{Timeline interface and impact graphs.}
    (a) An example timeline showing the presence of three intersectional identities (\textit{Male White, Male Black or African American}, and \textit{Female White}) in the Movie \textit{The Amazing Spiderman 2} (2014).
    Male White characters are present throughout the storyline.
    Further, there appears to be a lack of interactions (identified by the grey encompassing x-axis lines) between Male Black or African-American and Female White characters except for a few aberrant ones.
    However, both groups have interactions with Male White characters.
    The orange bars show one such interaction between Male White and Female White characters.
    (b-e) Example impact graphs for the characters Peter Parker, Harry Osborne, Max Dillon, and Gwen Stacy.}
    \label{fig:timeline2}
\end{figure*}

Based on the identities defined by a user in the Characters and Demographics Panel, the timeline can also show the presence of different identities in the timeline (Figure~\ref{fig:timeline}c). Using the \textit{Show} dropdown in the timeline~(Figure~\ref{fig:timeline}a), a user can choose to show the timeline for characters or demographics.
A user can also choose to show the presence of intersectional identities in the timeline.
For example, Figure~\ref{fig:timeline2}a presents a timeline showing the presence of intersectional groups in the Movie \textit{The Amazing Spiderman-2}. 
Note that for brevity, we do not show the Characters and Demographics Panel for this movie.
For the relevant Characters and Demographics Panel, please refer to the supplemental video.
We also note that the race assigned for the characters of this movie is based on the perception of the authors and may be different in reality.

In summary, the goal of the timeline visualization is to facilitate writers investigating agency for characters and social identities (\textbf{DG2}).
Since the task is subjective to a story and a writer's perspective, we intended to expose potential gaps in the story and let a writer explore and navigate the story easily (\textbf{DG1}).
We decided that visualizing the mentions of the characters in a timeline can be a starting point for this task and will allow users to see gaps easily.
The user can then use the timeline to further investigate any part of the story.
To aid this process, other visual components are also connected to the timeline, as described below. 

\paragraph{Impact Graph}

The timeline visualization primarily shows the presence of entities in a story.
While a user can identify the interactions between entities by observing overlaps in their presence in the x-axis, the timeline does not give a definitive answer to how an entity interacted with the other entities.
Additionally, multiple entities can be mentioned in a passage of a story; however, that does not necessarily mean they interacted with each other.

To overcome this shortcoming, we introduced the \textit{Impact Graph}, a force-directed network visualization that shows interactions between entities (\textbf{DG2}).
We consider an interaction between two entities if they are mentioned together in a sentence.
The impact graph for an entity is available whenever a user clicks on a y-axis label in the timeline. 

Figure~\ref{fig:timeline2}b-d show impact graphs for Peter Parker, Harry Osborne, Max Dillon, and Gwen Stacy from the movie \textit{The Amazing Spiderman 2}.
The selected entity is placed at the center of the network.
The edge widths represent the strength of the interactions.
For reducing clutter, only edges with at least five interactions are shown.
We observe that the impact graph for Gwen Stacy is much smaller than the other three male characters. 
Further, Gwen Stacy has significant interactions with Peter Parker only.

\paragraph{Word Zones}

Word cloud is a popular visual representation for showing a collection of words.
They have aesthetic value to lay users and are fun, and engaging~\cite{hearst2019evaluation, viegas2009participatory}.
In contrast, researchers have shown that they are not well-suited for analytic tasks such as finding a word and comparing the frequencies of words~\cite{hearst2019evaluation}.
To balance the utility and aesthetic value of word clouds, Hearst et al.~\cite{hearst2019evaluation} proposed Word Zones, a variation of word clouds, where words are grouped based on predefined labels/categories.
Since our users will most likely be non-experts in terms of visualization expertise, we decided to use Word Zones, thus opting for a visualization that is expected to be well-known to the writers as well as has better representation for analytic purposes (\textbf{DG6}).

A writer can add an entity for seeing words used with the entity in the word zone visualization whenever a user clicks on a y-axis label in the timeline (\textbf{DG3}). A writer can add as many entities as they want in the word zone. A user can control the number of words to show for each entity in the word zone using a slider. Users also have a dropdown to see relevant adjectives, verbs, or both. The verbs and adjectives for characters are extracted using dependency parsing. They are then aggregated for the social identities. Overall, this mechanism helps in answering questions such as: \textit{How the female characters were described in the story?} \textit{What were the actions of female characters in the story?}~(\textbf{DG3}).

 We considered each entity (a character or identity) as a document and the full story as a corpus of documents (entities). Based on that, the weight of a word ($w$) for an entity ($e$) is calculated as:

\begin{equation}
\label{eq:1}
    weight(w, e) = tf(w,e) * (1/df(w)) 
\end{equation}

where $tf(w,e)$ is the frequency of $w$ in $e$
and $df(w)$ is the frequency of $w$ in the whole story. It is essentially a normalized version of \textit{tf-idf}, popularly used for filtering out common and stop words, finding words of interest. However, \textit{tf-idf} is often applied on large corpora. In our case, the number of entities is limited to at most a few hundred. We also consider adjectives and verbs only. Thus, we opted for a simple normalization. 

\begin{figure}[t]
    \centering
    \includegraphics[width=0.95\columnwidth]{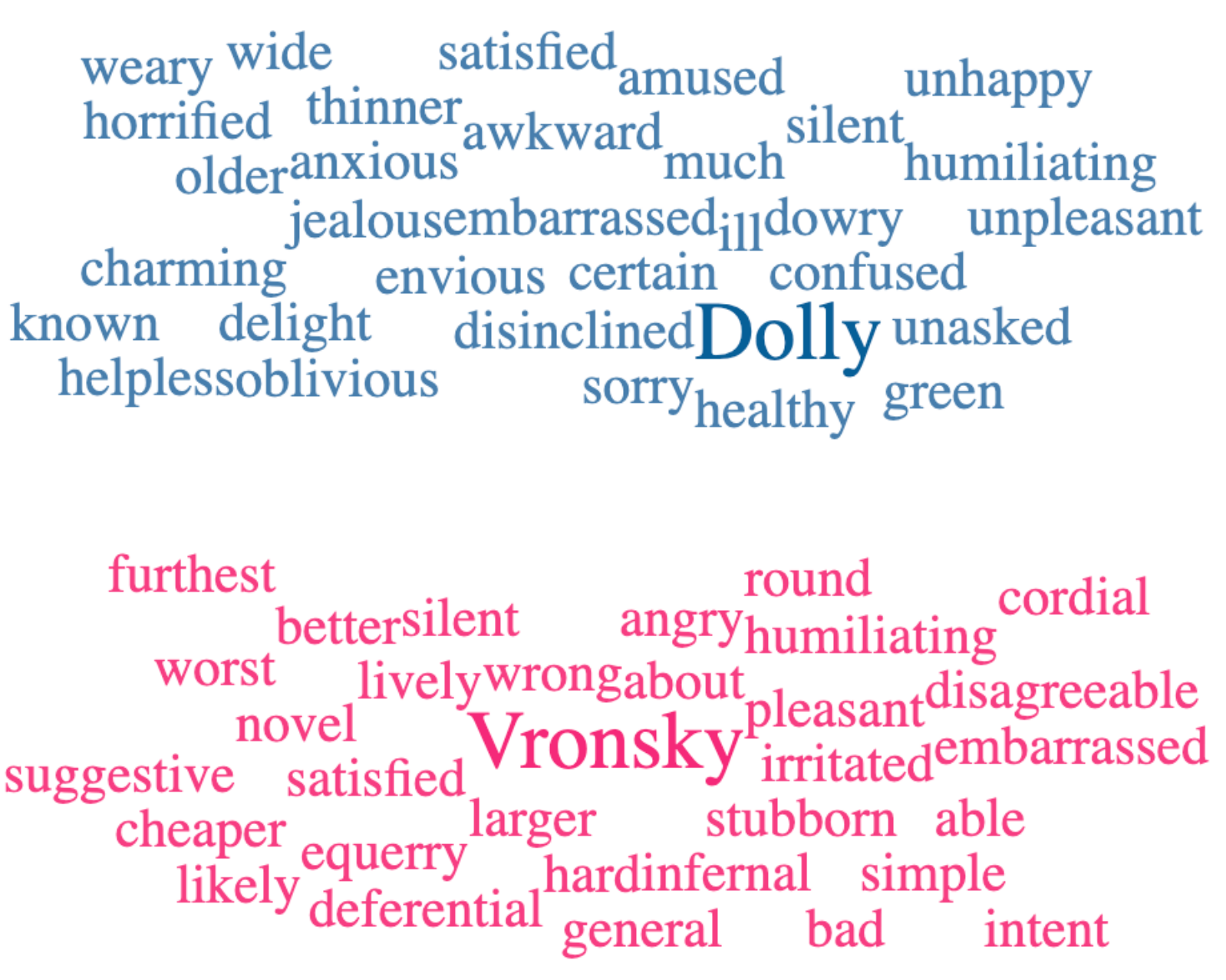}
    \caption{Word Zone representation for Dolly and Vronsky from \textit{Anna Karenina} (1877) by Leo Tolstoy, publicly available under Project Gutenberg~\cite{anna_kerreina}.}
    \Description{A picture showing examples of Word Zone visualization.}
    \label{fig:word_zone}
\end{figure}

Figure~\ref{fig:word_zone} shows an example word zone showing adjective used for Dolly (a female character), and Vronsky (a male character) from \textit{Anna Karenina} (1877) by Leo Tolstoy. Note the words such as ``charming'', ``envious'', ``jealous'', ``helpless'', ``oblivious'' and others in Dolly's word zone.

\begin{figure*}
    \centering
    \includegraphics[width = 0.9\textwidth]{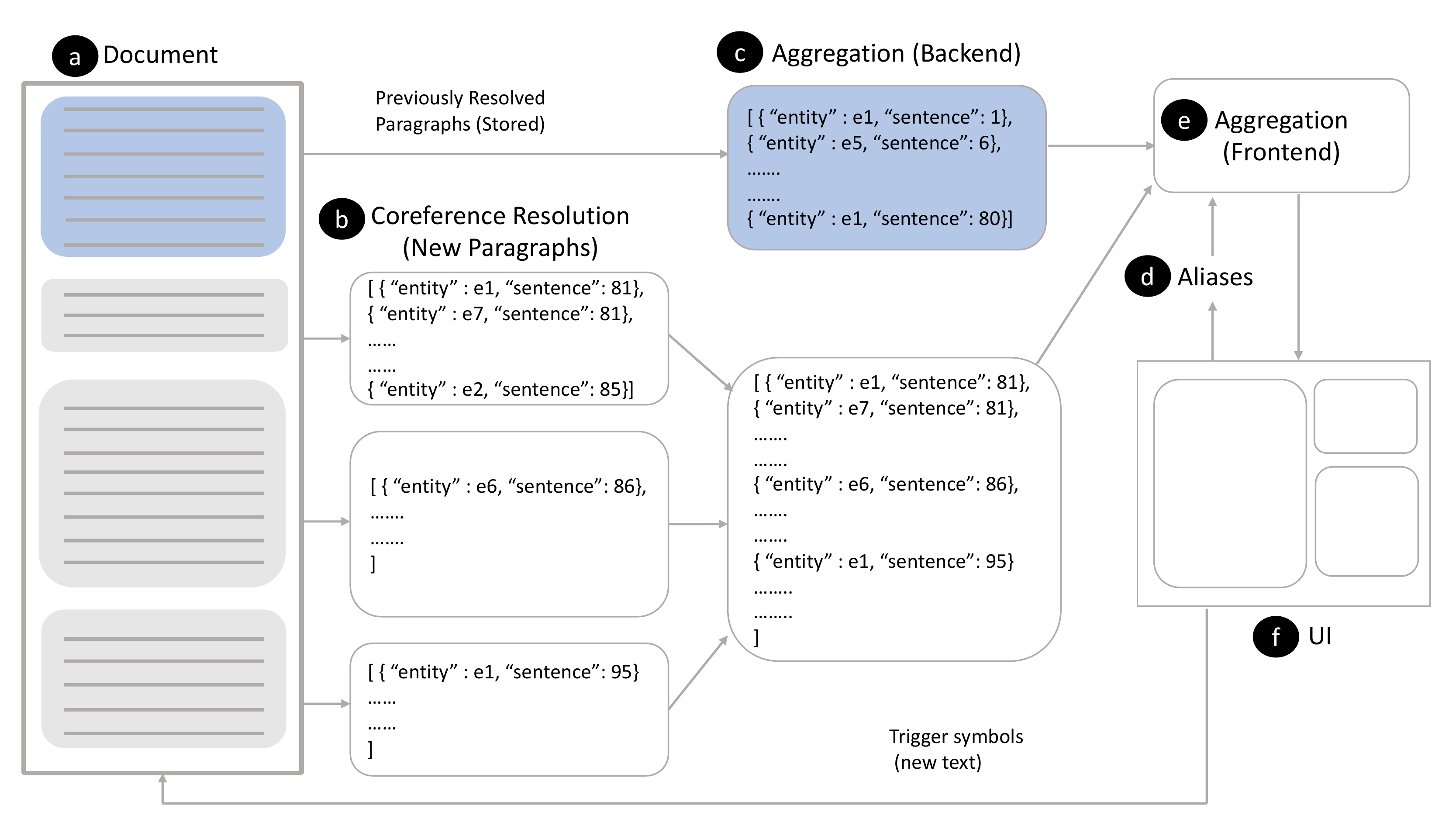}
    \Description{A picture showing analysis pipeline for the DVP tool.}
    \caption{\textbf{Example workflow of DVP.}
    (a) The system detects three new paragraphs in the text editor.
    The blue box represents the previously written text while the grey boxes represent the newly added paragraphs.
    (b) The system runs co-reference resolution and entity extraction independently on the newly added paragraphs.
    (c) They are then aggregated together with the stored entity information of the previous text.
    (e) A second aggregation on the front end is applied based on the aliases (d) identified by the user.
    (f) Finally the JSON formatted data is fed to the UI. The UI listens for the trigger symbols such as ``.'', ``?'', ``!'' or copy/paste event in the text editor for updating the UI. }
    \label{fig:implementation}
\end{figure*}

\subsection{Information Extraction from Text}

We used NeuralCoref~\cite{neuralcoref} along with Spacy~\cite{spacy} for extracting named entities, and their mentions in a document (Coreference Resolution~\cite{jurafskyspeech}).
Both packages are considered state-of-the-art and widely used in different applications.
However, coreference resolution for long documents such as a book can take a lot of time and system memory \cite{toshniwal2020learning}.
This could hinder the usability of the tool as our tool is targeted as a writing tool that provides instant feedback to writers as they write (\textbf{DG6}).

To increase the scalability of computation and reduce processing time, we adopted a divide-and-conquer method and run the coreference resolution model paragraph-wise, instead of the full document together.
We noticed that it is unlikely that a character would be referenced with a pronoun without the actual noun of the character in a new paragraph.
We reached out to the participants from the interview studies and confirmed that this is the usual case.
Figure~\ref{fig:implementation} presents an example scenario of our method.
On detecting a new trigger symbol such as ``.'', ``?'', ``!'' or a copy/paste event, the UI sends the document and the current Delta~\cite{delta} object to the server.
The Delta object contains information about how many and what characters (symbols) are inserted, deleted, or retained since the last update.

Upon receiving the document, the server splits the document into paragraphs. The paragraphs are identified by double newlines or ``\textbackslash n\textbackslash n''.
Based on the Delta object, the server then determines which paragraphs are retained (same as the previous update), and which paragraphs have new contents (insert or delete).
In Figure~\ref{fig:implementation}, the server detects three new paragraphs (grey boxes).
The server then runs entity recognition and coreference resolution models on the newly detected paragraphs.
Although our current implementation processes the paragraphs sequentially, they can be processed in a parallel fashion since the paragraphs are mutually independent.
The information extracted from the newly added paragraphs is then aggregated together with the stored information of previously processed paragraphs.
The client-side then performs another aggregation to combine aliases.


\subsection{Implementation Notes}

DVP is a web-based writing tool.
We used Python as the back-end,  JavaScript as the front-end language, and D3~\cite{bostock2011d3} for interactive visualization.
Semantic-UI and Bootstrap were used for styling various visual objects. 
 We used Spacy~\cite{spacy} and NeuralCoref~\cite{neuralcoref} for all the NLP tasks. The tool is available here: \url{https://github.com/tonmoycsedu/DramatVis-Personae}.
 


\section{Evaluation}

We evaluated DVP in two parts.
First, we conducted think-aloud sessions with three writers to understand the potential of DVP as a writing tool and identify potential usability issues.
We then conducted a user study with 11 writers to identify the benefits of DVP in comparison to a simple text editor.
We describe each study in turn below.

\subsection{Think-aloud Sessions}

For this study, we invited 3 writers to test our tool.
Two of them participated in our formative study (W2 and W5 from Table~\ref{tab:participant}).
We refer to them as W1 and W2 in the following section.
The other writer (W3) did not participate in the formative study.
W3's self-reported demographics are male, white, 44 years of age, with more than 20 years of writing experience.

Before the start of the study sessions, we emailed each writer independently asking whether they would like to write a new short story using our tool, or they would like to analyze stories that have already been written by them.
W1 and W2 wished to analyze previously written stories while W3 wished to write a story using our tool.
We asked writers that their stories should have at least three significant (named) characters and any number of supporting characters, at least two different scenes, and some dialogue between the characters.
Additionally, we encouraged W3 to ideate and take notes on their story and characters, but asked not to begin the actual writing process until the session.

One author of this paper administrated the sessions.
After consent, the administrator demonstrated the tool with a sample story. We encouraged participants to ask questions at this stage.
Once participants felt comfortable with the tool, we asked them to analyze a story (W1 and W2) or generate a new short story (W3).
While using the interface, participants thought aloud and conversed with the administrator regularly. Each session ended with an exit interview.
The sessions with W1 and W2 took around 1 hour while the session with W3 lasted around 2 hours.
All sessions were conducted using Zoom video conferencing and were audio and video recorded for post-session analysis.
At the end of the interview, each participant received a \$15/hour worth Amazon gift card.
The study administrator and another author of this paper independently analyzed the recorded videos, and study notes and then met together to sort the feedback from the writers and our observations into the following five thematic categories:

\subsubsection{Intuitive and Easy-to-use}

All participants found the interface to be easy and intuitive.
Overall, the design, especially the incorporation of visualizations in a writing tool, intrigued participants.
Initially, W2 and W3 were slightly confused about the timeline representation but quickly became accustomed to the timeline once they started interacting with the tiles.
Participants appreciated that the visualization and text editor were connected, and that writers can easily use the timeline to go to any part of a story.
We did not have to provide any explanation for the word zones; participants were already familiar with them.
Finally, while it took participants a few minutes to validate the characters, merge aliases, and then assign social identities to the characters, all participants were enthusiastic about this process.
W3 suggested a few stylistic changes such as auto-indenting a new paragraph and adding a thesaurus.

\subsubsection{Effect of Assigning Social Identities to Characters}

The mere presence of an interface where writers can assign social identities to characters had an effect on the writers.
W3 said: \textit{``This is an important feature for me as I usually plan very little before writing, and this helped me incorporate planning early into the writing and made me think how the characters should be represented in the story.''}
W2 mentioned that assigning social identities to characters could be a good practice for writers, especially when characters do not share identities with the writer.
This will help them think about the characters from the beginning.

\subsubsection{Usage Patterns}

All participants mentioned that DVP does not obstruct their creative process.
They appreciated that by using DVP they can explore biases when they want to and concentrate on writing or reading other times.
For example, W3 used the ``Hide'' option to hide the visualizations while writing and only checked them periodically.
W2 also suggested that they will probably check the visualizations after drafting a scene or section.

Both W1 and W2 extensively used the word zone by comparing words used for characters and social identities.
W2 suggested that it would be useful to quickly find the characters that have significantly different adjectives or verbs.

\subsubsection{Errors in Identifying Character Mentions}

DVP is powered by NLP toolkits and their capabilities to extract character names and identify character mentions throughout the story.
The pipeline may sometimes not recognize characters or their relevant mentions.
We noticed such cases created confusion among writers.
One of the animal characters from W2's story was not identified by our tool.
W2 asked whether our tool can identify unusual names since their sci-fi stories often have names that are not common.
Similarly, W1 noticed the tool missing out a few mentions of a character in the timeline.
While we expect the mitigation of such problems as NLP techniques become increasingly powerful, an alternative would be to allow writers to highlight name entities in the text editor and manually add them for tracking.

\subsubsection{Potentials and Impact}

At exit interviews, participants discussed several use cases and the impacts of DVP.
All participants suggested that DVP would be helpful to any creative writer.
They admired its support for different forms of writing (e.g., screenplays, fiction, non-fiction) which makes it usable for a wide range of creative materials.
W1 teaches novice journalists on how to write critiques, and was interested in using the tool to show examples of correct and biased representation to the students.
Further, W1 thought novice journalists can use the DVP tool as a learning tool and editors can use it to quickly check drafts.

\begin{figure*}
    \centering
    \subfloat[]{\includegraphics[width = 0.45\textwidth]{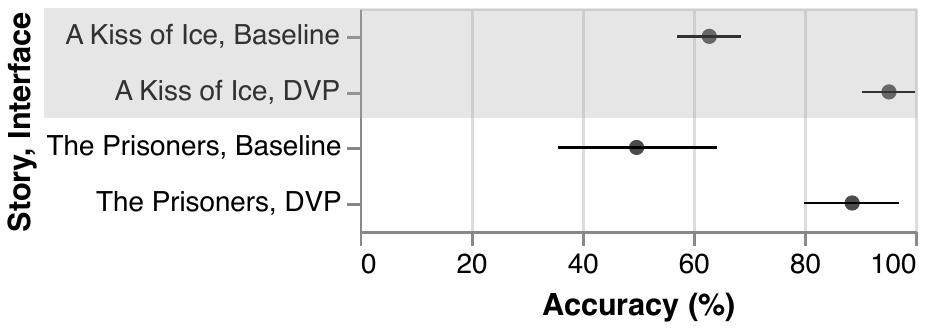}\label{fig:accuracy}}
    \quad \quad
    \subfloat[]{\includegraphics[width = 0.45\textwidth]{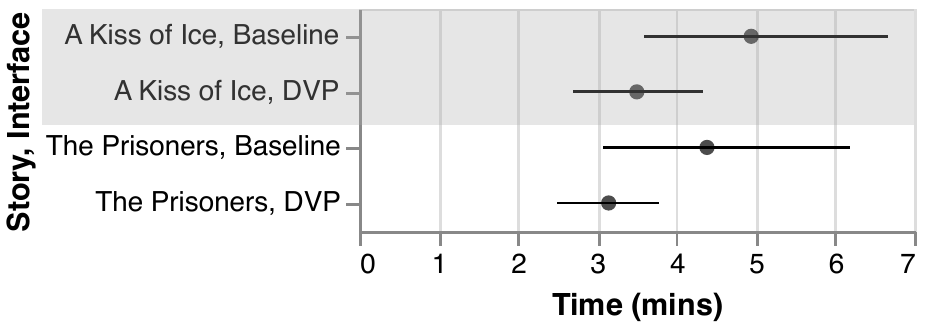}\label{fig:time}}
    \hfill
    \subfloat[]{\includegraphics[width = 0.45\textwidth]{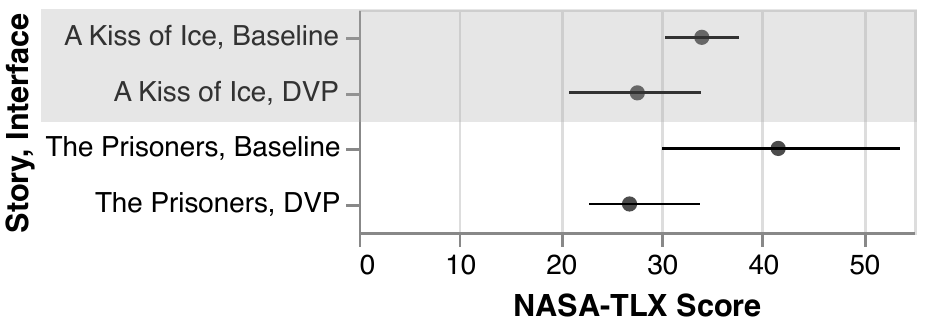}\label{fig:nasatlx}}
    \quad \quad
    \subfloat[]{\includegraphics[width = 0.45\textwidth]{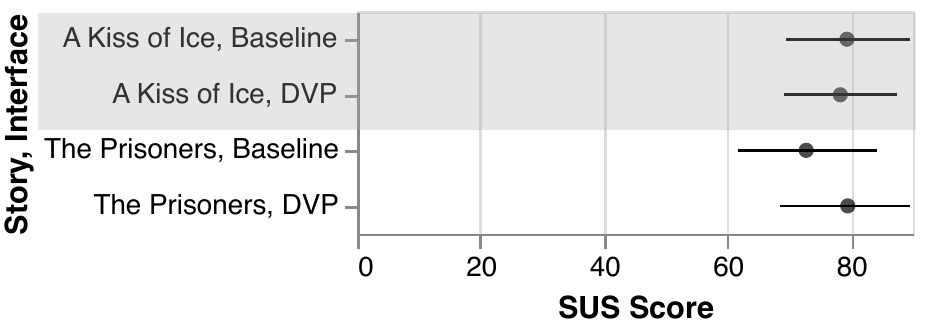}\label{fig:sus}}
    \Description{A picture showing results from the user study.}
    \caption{\textbf{Study Results.}
    (a) Participants' accuracy in answering the study questions; (b) Task completion time in minutes; (c) NASA-TLX (perceived workload) score; and (d) SUS score (usability ratings).
    Error bars show 95\% confidence intervals (CIs).}
    \label{fig:my_label}
\end{figure*}

\subsection{User Study}


We conducted a user study to measure the effectiveness of DVP in identification of biases. We designed a repeated-measures within-subject study with two interfaces: \textbf{I1. Baseline}, presented a simplified version of DVP, featuring only the text editor (Figure~\ref{fig:teaser}a) and demographic information of the characters (Figure~\ref{fig:teaser}b); and \textbf{I2. DVP Tool}, presented our interface with all features. We provide a screenshot of the baseline in the supplement.




Before conducting this study, based on the feedback from the think-aloud sessions, we made the following three refinements to our prototype.
First, we added a feature for users to manually highlight a character in the text editor and add it to the Characters and Demographics Panel for tracking.
This ensures that any character missed by the NLP pipeline can be supported in the system.
Second, the study session suggested that while word zone provides an easy way to investigate adjective and verbs, the search space to explore all the combination of entities can be large.
To aid writers in this process, we introduced a word embedding based approach to find potential candidate pairs of entities for investigation.
Let $\vec{g1}$ be the mean vector representing the words from entity $ent1$ and $\vec{g2}$ be the mean vector representing the words from entity $ent2$.
Then, the cosine distance between $\vec{g1}$ and $\vec{g2}$ can indicate how different the words from $ent1$ and $ent2$ are. This method closely matches the \textit{relative norm difference}~\cite{100years, ghai2021wordbias}, previously used to quantify gender bias in word embeddings.
The top 10 pairs of entities sorted by their cosine distance  are fed to the interface as a notification in the word zone control panel.
Finally, we added a few convenience features such as auto indent and word count in the system.

\subsubsection{Participants}

We recruited 11 participants through local mailing lists, university mailing lists, and public posts on Facebook groups and Twitter.  
Our inclusion criteria included prior experience with any forms of creative writing (fiction, non-fiction, short story, screenplays, etc).
We did not require participants to have published articles in their portfolios.
The participants varied in age from 21 to 45 (mean 26.5, s.d.\ 4.32), gender (5 males, 6 females), and race (6 Asian, 4 White, and 1 African American or Black).
All participants had prior experience with writing creative materials and using a text editor.
None of them participated in any of our prior studies.
Participants received \$15 per hour worth of Amazon gift cards for their participation.

\subsubsection{Data: Short Stories}



We used two newly written short stories for the study. One of the authors of this paper who had published fiction in their portfolio wrote one short story (``A Kiss of Ice'') for the study.
The other story (``The Prisoners'') was written by a professional writer, hired from \url{Fiverr.com}.
We compensated the professional writer with \$50.
As requirements, we asked both writers to write a short story that includes at least 4-5 characters and 3-4 different intersectional identities with a significant amount of interaction (having conversations, or participating in a scene together) between them. We also requested them to introduce a few stereotypes in the stories, hoping they will incite meaningful discussion in the exit interviews.
The two stories and social identities of the characters are provided in the supplement. 

\subsubsection{Study Questions}

DVP is designed to support the identification of two types of biases:
(1) lack of agency for characters;
and (2) stereotypes in describing characters.
Thus, we wanted to test how effective DVP would be in detecting these biases.
To do so, we designed 7 questions for each story that either fall into the first or second bias, or both.
A few examples of questions are
\textit{``Which of the following characters has the least presence in the story?
Which of the following social identities has the least presence in the story?
Which of the following two characters have the least interactions (having conversations, or participating in a scene together) with each other in the story?
How would you categorize how Female characters were portrayed in the story?''}

We provided multiple choices for each question.
The questions were similar for both stories although they were slightly modified to adjust to the nature of the stories. Two authors of this paper separately and manually analyzed the stories to get the correct answers and verified them by discussing with each other. The correct answers were also validated by the writers of the stories.
All questions are provided in the supplement.
Note that DVP is capable of supporting more demanding tasks such as sorting presence of every character; however, we intentionally did not include these tasks as they might be exhausting and overbearing to perform in a plain text editor (our baseline interface).




\subsubsection{Procedure}

Similar to the think-aloud sessions, we conducted the study sessions via Zoom. The study participants were provided a url where DVP was hosted. They interacted with the system via their own web browser.
A study session began with the participant signing a consent form.
Following this, the participants were introduced to the assigned first condition.
All participants had prior experience in writing and reading in a simple text editor (I1) so they did not require any training for this interface.

For DVP (I2), we provided a brief description of the interface and a demo showing its different features.
The participants then interacted with the system (with a training story), during which they were encouraged to ask questions until they were comfortable.

Each participant was then given a short story for the assigned first interface and asked to read the story. 
On average, it took participants 10 minutes (min = 5 minutes, max = 18 minutes) to read the short story.
After participants finished reading the story, we provided them with the task list (questions) for the first interface and asked them to answer the questions using the interface within 15 minutes. The questions were presented in random order to the participants. We provided them verbal clarification for the questions, if needed.
At the end of the first interface, we administrated the NASA-TLX~\cite{hart2006nasa} questionnaire to measure the participants' perceived workload, and SUS questionnaire to measure user experience and usability of that interface.
The same process was carried out for the second interface.
Each session lasted around 1.5 hours and ended with an exit interview.



\subsubsection{Study Design Rationale}

Our research objective from this study was to determine whether DVP can actually help writers identify biases. 
To evaluate that, we contemplated different study tasks and story types described as follows:

\paragraph{Task.} The ideal scenario would perhaps be to ask writers to write a long and complicated story using DVP since that is the primary use case of DVP.
However, given the sensitive nature of bias, it is unlikely that writers would be comfortable analyzing their work critically and answering questions related to bias in the presence of a study administrator.
Additionally, writing a story, even a short one, takes significant planning and effort, and is likely to be a very long and involved process.

An alternative would be to ask writers to use our tool on their own and provide feedback periodically.
While that might lead to insightful feedback, the process is uncertain and it will require a long-term commitment from a group of writers.
We felt that this was impractical at this stage of the research project.
Thus, we decided to avoid asking writers to write new stories in this study.

\paragraph{Story Type.} Since we decided to avoid the writing task, we needed already written stories for the study. We considered using stories written by the writers themselves or freely available stories online.
In both cases, participants may have already read and analyzed the stories.
Thus, using them may not lead us to isolate the effect of DVP.
Additionally, we needed to know the social identities of the characters, which are typically not explicitly provided in published articles. 
Considering these factors, we decided to use newly written stories for the study.

\subsubsection{Results}

To measure performance, we calculated participants' accuracy in answering the questions.
We also measured task completion time, perceived workload (NASA-TLX), and usability (SUS).
Following recent guidelines for statistical analysis in HCI~\cite{dragicevic2016fair}, we intentionally avoided traditional null-hypothesis statistical testing (NHST)
in favor of estimation methods to derive 95\% confidence intervals (CIs) for all measures.
More specifically, we employed non-parametric bootstrapping~\cite{efron1992bootstrap} with $R = 1,000$ iterations.
We also report mean differences as sample effect size and Cohen's $d$ as a standardized measure of effect size~\cite{cohen1988statistical}.



\paragraph{Accuracy.}
%
Figure~\ref{fig:accuracy} presents participants' accuracy in answering the questions. 
Overall, participants were more accurate when answering the questions using DVP.
For the ``A Kiss of Ice'' story, participants were more accurate on average by 32.49 ($d$=4.27) percentage points using DVP.
Similarly, for ``The Prisoners'' story, participants were more accurate on average by 38.25 ($d$=2.31) percentage points using DVP.
The very large effect sizes suggest strong practical difference.
%
%
We noticed participants performed slightly better for the story ``A Kiss of Ice'' than ``The Prisoners'' using both interfaces.
We anticipated this result as ``The Prisoners'' featured frequent interactions between different characters and social identities which raised the difficulty level for answering the questions.

\paragraph{Task Completion Time.} 
%
Figure~\ref{fig:time} presents participants' task completion time.
The completion time was measured by calculating number of minutes participants took to answer the study questions, excluding the time to read the stories.
Unsurprisingly, participants spent less time to answer the questions using DVP.
For ``A Kiss of Ice,'' participants spent on average 1.37 ($d$=0.88) minutes less using DVP.
Similarly, for ``The Prisoners,'' participants spent on average 1.24 ($d$=0.72) minutes less using DVP.


\paragraph{Perceived Workload (NASA-TLX).}
%
Figure~\ref{fig:nasatlx} presents NASA-TLX scores, average of the raw ratings provided by the participants on six commonly used workload measures: mental demand, physical demand, temporal demand, effort, performance, and frustration~\cite{hart2006nasa}.
DVP reduced workloads of participants in answering the questions.
For ``A Kiss of Ice,'' on average the perceived workload of the participants were 5.66 ($d$=0.84) points less using DVP. 
Similarly, for ``The Prisoners'' on average the perceived workload of the participants was 14.79 ($d$=1.08) points less using DVP.


\paragraph{Usability (SUS Score).}
%
Participants rated both condition similarly in terms of usability.
Figure~\ref{fig:sus} presents SUS scores calculated from the 10 individual SUS questions.
As observed in Figure~\ref{fig:sus}
, DVP did not warrant any major usability issues.
\vspace{6pt}

\paragraph{Qualitative Feedback.}
Participants mentioned that DVP adds a new dimension to reading stories (P2, P4, P7-9).
The demographic information of the characters and visualization created a situational awareness among them.
For example, P4 said \textit{``Normally, I do not think about the identities of the characters while reading, beyond what can be inferred from the names, and description of the characters.
The visualization certainly created a new dimension to my reading and I can imagine the characters in new ways that I would not do usually.''}
We noticed several participants used the default search and find feature of browsers when answering the questions using the simple text editor.
In the exit interviews, several participants mentioned that they used the search and find feature as a way to quickly find relevant information for answering the questions.
However, it was not effective in answering the questions as they needed to mentally analyze the search results along with the demographic information which might not be directly encoded as key words in the story.
In contrast, the visualization in DVP supported them in answering the questions by providing summarized information.
Participants also saw value in DVP as a writing tool.
They mentioned that the tool will be most useful for writing long stories with many inter-connected characters and subplots (P1-5, P7, P9).
Participants mentioned several use cases where DVP may be helpful.
For example, P1 said \textit{``I think this tool has a clear application for writing adapted screenplays.
When I write an adapted screenplay, I constantly try to evaluate my writing with the representation in the original book.
The tool can be helpful in such scenarios.'' } 


\section{Discussion, Limitations, and Future Work}

In this section, we discuss potential implications and limitations learned from the development and evaluation of DVP.

\paragraph{Long Term Deployment of DVP.}
The think-aloud sessions and user study show promise for DVP as a new medium for writing and critical reading.
Even for short stories that were used in the user study, DVP has made the bias identification process much easier for writers.
Further, all participants were enthusiastic about the tool and thought it would be useful for them.
To evaluate DVP to its full extent, our future work will concentrate on a long term deployment of DVP with writers (e.g., writing a full-length story using DVP).

\paragraph{Impact on Bias in Creative Writing.}
Bias in creative writing is ubiquitous, as described in Section~\ref{sec:story-bias}.
Despite our success with DVP, we believe it is important to acknowledge that our work will not solve systematic or infrastructural discrimination that exists in the publishing community. 
For example, there is still a lack of writers from marginalized groups in the current writer's community~\cite{gd_equity2021}. 
We consider DVP as a part of the larger movement against biases in creative writing, a probe for reducing biases, and a catalyst for future efforts in this direction.
We hope its adaptability to different forms of writing (novels, children's books, short stories) will attract diverse writers. 




\paragraph{Balancing Automation and Artistic Freedom/Agency.}
From the beginning of this project, we were vigilant about the artistic freedom of writers, and how we can balance agency with automation.
We understand writers may want to intentionally write about discrimination and biases against a social group, and we do not want to see the tool preventing them in this process.
That is one of the reasons why we did not use any automation in correcting biases and stereotypes.



It is also worth noting here that our intention was not to replace critique groups or sensitivity readers.
We believe it is important to have subjective feedback about creative writing, especially from relevant marginalized groups.
Our intention was to help writers during their writing process by offering another set of eyes---albeit electronic ones.

\paragraph{Design Implications for Human-centered AI.}
AI systems can inherit harmful biases and stereotypes.
These biases can impact social groups disparately, especially when used as a decision-making platform for critical resources ~\cite{scheuerman2019computers, angwin2016machine, ghai2022cascaded}.
These systems may also lack inclusivity (e.g., lack of supports for non-binary identities).
These limitations motivated several design decisions of our system.

First, based on DG5, we made the \textit{demographics} dropdowns unconstrained so that a writer could add any social identity required for the story.
Second, we provided several functions in the interface for the writers to validate (e.g., merge, delete) the results returned by the NLP pipeline, a safety check against potential biases in coreference resolution~\cite{zhao2018gender}, and dependency parsing~\cite{garimella2019women}.
Additionally, a writer can interactively add an entity for tracking in the case the entity was not recognized by the NLP pipeline.
Thus, our work promotes human-centered AI for creative writing.

\paragraph{Limitations of the Underlying AI.}
NLP methods employed in this paper are highly accurate; however, they are not error free.
In particular, it is worth noting that different NLP models like coreference resolution, named entity recognition might exhibit biases based on social identities like gender~\cite{zhao2018gender, mehrabi2020man}.
This might result in skewed performance against minority groups~\cite{mehrabi2020man, mishra2020assessing}. 
Moreover, the NLP models used in this work were trained on a set of news articles, weblogs, etc.
Such data might differ from literary texts like novels books, etc. as they might contain longer sentences, more sophisticated language, etc~\cite{ rosiger2018towards}.
Although we did not observe any such issues in our experiments, future work might employ fairness-aware NLP models that are trained on domain-specific datasets like books, novels, etc. for better performance.   


Finally, the design and development process of this tool has been influenced by English-speaking study participants and NLP models trained over English language corpus.
Hence, our tool can currently only support the English language.
Having said that, we know that social biases based on gender, race, etc, transcend societies and languages~\cite{fr_es, hindi}.
We believe that ideas put forward in this work will help facilitate the development of similar tools for other languages as well.
Future work might support others languages like French, Chinese, etc. by incorporating NLP models trained over different language corpora. 

\section{Conclusion}

In this paper, we have presented an in-depth case study on supporting creative writing and combating implicit bias in fiction using interactive technologies, data visualization, and NLP.
In particular, we have reported on an interview study involving 9 creative writers where we asked them about their process, how they navigate harmful stereotypes, and how they think tool support could help in this work. 
Based on these interviews, we designed \textsc{DramatVis Personae}{} (DVP), a visual analytics tool using NLP to visualize characters, their demographics, and their story arcs in an effort to mitigate implicit bias.
The tool can be used both in an online manner while writing a story, as well as offline during the analysis of an already written story. 


We believe that our work here suggests many interesting future avenues of research. 
For one thing, while creative writing is a notoriously individual and idiosyncratic process, and while the human touch is vital to true art, our moderate success with \textsc{DramatVis Personae}{} points to possible ways to augment this human touch to improve even such famously crooked processes.
In particular, we think that our work shows how automatic machine eyes, while certainly less keen and discerning than human ones, can be helpful for certain applications such as mitigating bias if only because they---unlike human eyes---remain unblinking. 
We are not so foolish as to believe in the idea of an ``unbiased algorithm''---all algorithms are created by humans and thus intrinsically carry the biases of their creators---but we do believe in the virtue of training as many different lenses as possible on a creative artifact in the hope of uncovering yet another harmful stereotype or instance of implicit bias.
Thus, we tend to think that our work here is in no way indicative of an end of art, but rather a new beginning.

\begin{acks}
 This work was partially supported by the Doctoral Student Research Award from College of Information Studies, University of Maryland, College Park.
\end{acks}








\bibliographystyle{ACM-Reference-Format}
\bibliography{dramatvis}


\begin{thebibliography}{97}


\ifx \showCODEN    \undefined \def \showCODEN     #1{\unskip}     \fi
\ifx \showDOI      \undefined \def \showDOI       #1{#1}\fi
\ifx \showISBNx    \undefined \def \showISBNx     #1{\unskip}     \fi
\ifx \showISBNxiii \undefined \def \showISBNxiii  #1{\unskip}     \fi
\ifx \showISSN     \undefined \def \showISSN      #1{\unskip}     \fi
\ifx \showLCCN     \undefined \def \showLCCN      #1{\unskip}     \fi
\ifx \shownote     \undefined \def \shownote      #1{#1}          \fi
\ifx \showarticletitle \undefined \def \showarticletitle #1{#1}   \fi
\ifx \showURL      \undefined \def \showURL       {\relax}        \fi
\providecommand\bibfield[2]{#2}
\providecommand\bibinfo[2]{#2}
\providecommand\natexlab[1]{#1}
\providecommand\showeprint[2][]{arXiv:#2}

\bibitem[\protect\citeauthoryear{Abdul{-}Rahman, Lein, Coles, Maguire, Meyer,
  Wynne, Johnson, Trefethen, and Chen}{Abdul{-}Rahman et~al\mbox{.}}{2013}]%
        {DBLP:journals/cgf/Abdul-RahmanLCMMWJTC13}
\bibfield{author}{\bibinfo{person}{Alfie Abdul{-}Rahman},
  \bibinfo{person}{Julie Lein}, \bibinfo{person}{Katherine Coles},
  \bibinfo{person}{Eamonn Maguire}, \bibinfo{person}{Miriah~D. Meyer},
  \bibinfo{person}{Martin Wynne}, \bibinfo{person}{Chris~R. Johnson},
  \bibinfo{person}{Anne~E. Trefethen}, {and} \bibinfo{person}{Min Chen}.}
  \bibinfo{year}{2013}\natexlab{}.
\newblock \showarticletitle{Rule-based Visual Mappings - with a Case Study on
  Poetry Visualization}.
\newblock \bibinfo{journal}{\emph{Computer Graphics Forum}}
  \bibinfo{volume}{32}, \bibinfo{number}{3} (\bibinfo{year}{2013}),
  \bibinfo{pages}{381--390}.
\newblock
\urldef\tempurl%
\url{https://doi.org/10.1111/cgf.12125}
\showDOI{\tempurl}


\bibitem[\protect\citeauthoryear{Adukia, Eble, Harrison, Runesha, and
  Szasz}{Adukia et~al\mbox{.}}{2021}]%
        {adukia2021we}
\bibfield{author}{\bibinfo{person}{Anjali Adukia}, \bibinfo{person}{Alex Eble},
  \bibinfo{person}{Emileigh Harrison}, \bibinfo{person}{Hakizumwami~Birali
  Runesha}, {and} \bibinfo{person}{Teodora Szasz}.}
  \bibinfo{year}{2021}\natexlab{}.
\newblock \bibinfo{booktitle}{\emph{What we teach about race and gender:
  Representation in images and text of children’s books}}.
\newblock \bibinfo{type}{{T}echnical {R}eport}. \bibinfo{institution}{National
  Bureau of Economic Research}.
\newblock


\bibitem[\protect\citeauthoryear{Alencar, de~Oliveira, and Paulovich}{Alencar
  et~al\mbox{.}}{2012}]%
        {DBLP:journals/widm/AlencarOP12}
\bibfield{author}{\bibinfo{person}{Aretha~Barbosa Alencar},
  \bibinfo{person}{Maria Cristina~Ferreira de Oliveira}, {and}
  \bibinfo{person}{Fernando~Vieira Paulovich}.}
  \bibinfo{year}{2012}\natexlab{}.
\newblock \showarticletitle{Seeing beyond reading: a survey on visual text
  analytics}.
\newblock \bibinfo{journal}{\emph{Wiley Interdisciplinary Reviews: Data Mining
  and Knowledge Discovery}} \bibinfo{volume}{2}, \bibinfo{number}{6}
  (\bibinfo{year}{2012}), \bibinfo{pages}{476--492}.
\newblock
\urldef\tempurl%
\url{https://doi.org/10.1002/widm.1071}
\showDOI{\tempurl}


\bibitem[\protect\citeauthoryear{Angwin, Larson, Mattu, and Kirchner}{Angwin
  et~al\mbox{.}}{2016}]%
        {angwin2016machine}
\bibfield{author}{\bibinfo{person}{Julia Angwin}, \bibinfo{person}{Jeff
  Larson}, \bibinfo{person}{Surya Mattu}, {and} \bibinfo{person}{Lauren
  Kirchner}.} \bibinfo{year}{2016}\natexlab{}.
\newblock \showarticletitle{Machine bias}.
\newblock \bibinfo{journal}{\emph{ProPublica, May}}  \bibinfo{volume}{23}
  (\bibinfo{year}{2016}), \bibinfo{pages}{2016}.
\newblock


\bibitem[\protect\citeauthoryear{Arthur Quiller-Couch}{Arthur
  Quiller-Couch}{2016}]%
        {sleeping_beauty}
\bibfield{author}{\bibinfo{person}{Charles~Perrault Arthur Quiller-Couch}.}
  \bibinfo{year}{2016}\natexlab{}.
\newblock \bibinfo{title}{The Sleeping Beauty and other fairy tales}.
\newblock
\newblock
\urldef\tempurl%
\url{https://www.gutenberg.org/files/51275/51275-h/51275-h.htm}
\showURL{%
\tempurl}
\newblock
\shownote{Accessed: 08/30/2021.}


\bibitem[\protect\citeauthoryear{Beckett, Ellison, Barrett, and Shah}{Beckett
  et~al\mbox{.}}{2010}]%
        {beckett2010away}
\bibfield{author}{\bibinfo{person}{Angharad Beckett}, \bibinfo{person}{Nick
  Ellison}, \bibinfo{person}{Sam Barrett}, {and} \bibinfo{person}{Sonali
  Shah}.} \bibinfo{year}{2010}\natexlab{}.
\newblock \showarticletitle{`Away with the fairies?' Disability within
  primary-age children's literature}.
\newblock \bibinfo{journal}{\emph{Disability \& Society}} \bibinfo{volume}{25},
  \bibinfo{number}{3} (\bibinfo{year}{2010}), \bibinfo{pages}{373--386}.
\newblock


\bibitem[\protect\citeauthoryear{Bostock, Ogievetsky, and Heer}{Bostock
  et~al\mbox{.}}{2011}]%
        {bostock2011d3}
\bibfield{author}{\bibinfo{person}{Michael Bostock}, \bibinfo{person}{Vadim
  Ogievetsky}, {and} \bibinfo{person}{Jeffrey Heer}.}
  \bibinfo{year}{2011}\natexlab{}.
\newblock \showarticletitle{D$^3$ data-driven documents}.
\newblock \bibinfo{journal}{\emph{{{IEEE} Transactions on Visualization and
  Computer Graphics}}} \bibinfo{volume}{17}, \bibinfo{number}{12}
  (\bibinfo{year}{2011}), \bibinfo{pages}{2301--2309}.
\newblock


\bibitem[\protect\citeauthoryear{Braun and Clarke}{Braun and Clarke}{2006}]%
        {braun2006using}
\bibfield{author}{\bibinfo{person}{Virginia Braun} {and}
  \bibinfo{person}{Victoria Clarke}.} \bibinfo{year}{2006}\natexlab{}.
\newblock \showarticletitle{Using thematic analysis in psychology}.
\newblock \bibinfo{journal}{\emph{Qualitative Research in Psychology}}
  \bibinfo{volume}{3}, \bibinfo{number}{2} (\bibinfo{year}{2006}),
  \bibinfo{pages}{77--101}.
\newblock


\bibitem[\protect\citeauthoryear{Brown, Mann, Ryder, Subbiah, Kaplan, Dhariwal,
  Neelakantan, Shyam, Sastry, Askell, Agarwal, Herbert{-}Voss, Krueger,
  Henighan, Child, Ramesh, Ziegler, Wu, Winter, Hesse, Chen, Sigler, Litwin,
  Gray, Chess, Clark, Berner, McCandlish, Radford, Sutskever, and Amodei}{Brown
  et~al\mbox{.}}{2020}]%
        {brown2020language}
\bibfield{author}{\bibinfo{person}{Tom~B. Brown}, \bibinfo{person}{Benjamin
  Mann}, \bibinfo{person}{Nick Ryder}, \bibinfo{person}{Melanie Subbiah},
  \bibinfo{person}{Jared Kaplan}, \bibinfo{person}{Prafulla Dhariwal},
  \bibinfo{person}{Arvind Neelakantan}, \bibinfo{person}{Pranav Shyam},
  \bibinfo{person}{Girish Sastry}, \bibinfo{person}{Amanda Askell},
  \bibinfo{person}{Sandhini Agarwal}, \bibinfo{person}{Ariel Herbert{-}Voss},
  \bibinfo{person}{Gretchen Krueger}, \bibinfo{person}{Tom Henighan},
  \bibinfo{person}{Rewon Child}, \bibinfo{person}{Aditya Ramesh},
  \bibinfo{person}{Daniel~M. Ziegler}, \bibinfo{person}{Jeffrey Wu},
  \bibinfo{person}{Clemens Winter}, \bibinfo{person}{Christopher Hesse},
  \bibinfo{person}{Mark Chen}, \bibinfo{person}{Eric Sigler},
  \bibinfo{person}{Mateusz Litwin}, \bibinfo{person}{Scott Gray},
  \bibinfo{person}{Benjamin Chess}, \bibinfo{person}{Jack Clark},
  \bibinfo{person}{Christopher Berner}, \bibinfo{person}{Sam McCandlish},
  \bibinfo{person}{Alec Radford}, \bibinfo{person}{Ilya Sutskever}, {and}
  \bibinfo{person}{Dario Amodei}.} \bibinfo{year}{2020}\natexlab{}.
\newblock \showarticletitle{Language Models are Few-Shot Learners}.
\newblock \bibinfo{journal}{\emph{CoRR}}  \bibinfo{volume}{abs/2005.14165}
  (\bibinfo{year}{2020}), 75.
\newblock
\showeprint[arXiv]{2005.14165}
\urldef\tempurl%
\url{https://arxiv.org/abs/2005.14165}
\showURL{%
\tempurl}


\bibitem[\protect\citeauthoryear{Chaturvedi, Gannod, Mandell, Armstrong, and
  Hodgson}{Chaturvedi et~al\mbox{.}}{2012}]%
        {DBLP:conf/dihu/ChaturvediGMAH12}
\bibfield{author}{\bibinfo{person}{Manish Chaturvedi},
  \bibinfo{person}{Gerald~C. Gannod}, \bibinfo{person}{Laura Mandell},
  \bibinfo{person}{Helen Armstrong}, {and} \bibinfo{person}{Eric Hodgson}.}
  \bibinfo{year}{2012}\natexlab{}.
\newblock \showarticletitle{Myopia: {A} Visualization Tool in Support of Close
  Reading}. In \bibinfo{booktitle}{\emph{Proceedings of the Annual
  International Conference of the Alliance of Digital Humanities
  Organizations}}, \bibfield{editor}{\bibinfo{person}{Jan~Christoph Meister}}
  (Ed.). \bibinfo{publisher}{Hamburg University Press},
  \bibinfo{address}{Hamburg, Germany}, \bibinfo{pages}{148--150}.
\newblock


\bibitem[\protect\citeauthoryear{Chung, Kim, Yoo, Lee, Adar, and Chang}{Chung
  et~al\mbox{.}}{2022}]%
        {talebrush}
\bibfield{author}{\bibinfo{person}{John Joon~Young Chung},
  \bibinfo{person}{Wooseok Kim}, \bibinfo{person}{Kang~Min Yoo},
  \bibinfo{person}{Hwaran Lee}, \bibinfo{person}{Eytan Adar}, {and}
  \bibinfo{person}{Minsuk Chang}.} \bibinfo{year}{2022}\natexlab{}.
\newblock \showarticletitle{{TaleBrush}: Sketching Stories with Generative
  Pretrained Language Models}. In \bibinfo{booktitle}{\emph{Proceedings of the
  {ACM} Conference on Human Factors in Computing Systems}}.
  \bibinfo{publisher}{{ACM}}, \bibinfo{address}{New York, NY, USA},
  \bibinfo{pages}{to appear}.
\newblock


\bibitem[\protect\citeauthoryear{Clark, Ross, Tan, Ji, and Smith}{Clark
  et~al\mbox{.}}{2018}]%
        {clark2018creative}
\bibfield{author}{\bibinfo{person}{Elizabeth Clark},
  \bibinfo{person}{Anne~Spencer Ross}, \bibinfo{person}{Chenhao Tan},
  \bibinfo{person}{Yangfeng Ji}, {and} \bibinfo{person}{Noah~A Smith}.}
  \bibinfo{year}{2018}\natexlab{}.
\newblock \showarticletitle{Creative writing with a machine in the loop: Case
  studies on slogans and stories}. In \bibinfo{booktitle}{\emph{Proceedings of
  the ACM Conference on Intelligent User Interfaces}}.
  \bibinfo{publisher}{{ACM}}, \bibinfo{address}{New York, NY, USA},
  \bibinfo{pages}{329--340}.
\newblock


\bibitem[\protect\citeauthoryear{Coenen, Davis, Ippolito, Reif, and
  Yuan}{Coenen et~al\mbox{.}}{2021}]%
        {coenen2021wordcraft}
\bibfield{author}{\bibinfo{person}{Andy Coenen}, \bibinfo{person}{Luke Davis},
  \bibinfo{person}{Daphne Ippolito}, \bibinfo{person}{Emily Reif}, {and}
  \bibinfo{person}{Ann Yuan}.} \bibinfo{year}{2021}\natexlab{}.
\newblock \showarticletitle{Wordcraft: a Human-{AI} Collaborative Editor for
  Story Writing}.
\newblock \bibinfo{journal}{\emph{CoRR}}  \bibinfo{volume}{abs/2107.07430}
  (\bibinfo{year}{2021}), 7.
\newblock
\showeprint[arXiv]{2107.07430}
\urldef\tempurl%
\url{https://arxiv.org/abs/2107.07430}
\showURL{%
\tempurl}


\bibitem[\protect\citeauthoryear{Cohen}{Cohen}{1988}]%
        {cohen1988statistical}
\bibfield{author}{\bibinfo{person}{Jacob Cohen}.}
  \bibinfo{year}{1988}\natexlab{}.
\newblock \bibinfo{booktitle}{\emph{Statistical Power Analysis for the Social
  Sciences}}.
\newblock \bibinfo{publisher}{Erlbaum Associates}, \bibinfo{address}{Hillsdale,
  NJ, USA}.
\newblock


\bibitem[\protect\citeauthoryear{Correll, Park, Judd, and Wittenbrink}{Correll
  et~al\mbox{.}}{2007}]%
        {correll2007influence}
\bibfield{author}{\bibinfo{person}{Joshua Correll}, \bibinfo{person}{Bernadette
  Park}, \bibinfo{person}{Charles~M. Judd}, {and} \bibinfo{person}{Bernd
  Wittenbrink}.} \bibinfo{year}{2007}\natexlab{}.
\newblock \showarticletitle{The influence of stereotypes on decisions to
  shoot}.
\newblock \bibinfo{journal}{\emph{European Journal of Social Psychology}}
  \bibinfo{volume}{37}, \bibinfo{number}{6} (\bibinfo{year}{2007}),
  \bibinfo{pages}{1102--1117}.
\newblock


\bibitem[\protect\citeauthoryear{Correll, Witmore, and Gleicher}{Correll
  et~al\mbox{.}}{2011}]%
        {DBLP:journals/cgf/CorrellWG11}
\bibfield{author}{\bibinfo{person}{Michael Correll}, \bibinfo{person}{Michael
  Witmore}, {and} \bibinfo{person}{Michael Gleicher}.}
  \bibinfo{year}{2011}\natexlab{}.
\newblock \showarticletitle{Exploring Collections of Tagged Text for Literary
  Scholarship}.
\newblock \bibinfo{journal}{\emph{Computer Graphics Forum}}
  \bibinfo{volume}{30}, \bibinfo{number}{3} (\bibinfo{year}{2011}),
  \bibinfo{pages}{731--740}.
\newblock
\urldef\tempurl%
\url{https://doi.org/10.1111/j.1467-8659.2011.01922.x}
\showDOI{\tempurl}


\bibitem[\protect\citeauthoryear{Cui, Liu, Tan, Shi, Song, Gao, Qu, and
  Tong}{Cui et~al\mbox{.}}{2011}]%
        {Cui2011}
\bibfield{author}{\bibinfo{person}{Weiwei Cui}, \bibinfo{person}{Shixia Liu},
  \bibinfo{person}{Li Tan}, \bibinfo{person}{Conglei Shi},
  \bibinfo{person}{Yangqiu Song}, \bibinfo{person}{Zekai Gao},
  \bibinfo{person}{Huamin Qu}, {and} \bibinfo{person}{Xin Tong}.}
  \bibinfo{year}{2011}\natexlab{}.
\newblock \showarticletitle{{TextFlow}: Towards Better Understanding of
  Evolving Topics in Text}.
\newblock \bibinfo{journal}{\emph{{{IEEE} Transactions on Visualization and
  Computer Graphics}}} \bibinfo{volume}{17}, \bibinfo{number}{12}
  (\bibinfo{year}{2011}), \bibinfo{pages}{2412--2421}.
\newblock
\urldef\tempurl%
\url{https://doi.org/10.1109/TVCG.2011.239}
\showDOI{\tempurl}


\bibitem[\protect\citeauthoryear{Dekker, Kuhn, and van Erp}{Dekker
  et~al\mbox{.}}{2019}]%
        {dekker2019evaluating}
\bibfield{author}{\bibinfo{person}{Niels Dekker}, \bibinfo{person}{Tobias
  Kuhn}, {and} \bibinfo{person}{Marieke van Erp}.}
  \bibinfo{year}{2019}\natexlab{}.
\newblock \showarticletitle{Evaluating named entity recognition tools for
  extracting social networks from novels}.
\newblock \bibinfo{journal}{\emph{PeerJ Computer Science}}  \bibinfo{volume}{5}
  (\bibinfo{year}{2019}), \bibinfo{pages}{e189}.
\newblock


\bibitem[\protect\citeauthoryear{Dragicevic}{Dragicevic}{2016}]%
        {dragicevic2016fair}
\bibfield{author}{\bibinfo{person}{Pierre Dragicevic}.}
  \bibinfo{year}{2016}\natexlab{}.
\newblock \showarticletitle{Fair Statistical Communication in HCI}.
\newblock In \bibinfo{booktitle}{\emph{Modern Statistical Methods for HCI}}.
  \bibinfo{publisher}{Springer Publishing}, \bibinfo{address}{New York, NY,
  USA}, \bibinfo{pages}{291--330}.
\newblock


\bibitem[\protect\citeauthoryear{Efron}{Efron}{1979}]%
        {efron1992bootstrap}
\bibfield{author}{\bibinfo{person}{Bradley Efron}.}
  \bibinfo{year}{1979}\natexlab{}.
\newblock \showarticletitle{Bootstrap methods: another look at the jackknife}.
\newblock \bibinfo{journal}{\emph{The Annals of Statistics}}
  \bibinfo{volume}{7}, \bibinfo{number}{1} (\bibinfo{year}{1979}),
  \bibinfo{pages}{1--26}.
\newblock
\urldef\tempurl%
\url{https://www.jstor.org/stable/2958830}
\showURL{%
\tempurl}


\bibitem[\protect\citeauthoryear{Emons, Wester, and Scheepers}{Emons
  et~al\mbox{.}}{2010}]%
        {emons2010he}
\bibfield{author}{\bibinfo{person}{Pascale Emons}, \bibinfo{person}{Fred
  Wester}, {and} \bibinfo{person}{Peer Scheepers}.}
  \bibinfo{year}{2010}\natexlab{}.
\newblock \showarticletitle{“He Works Outside the Home; She Drinks Coffee and
  Does the Dishes” Gender Roles in Fiction Programs on Dutch Television}.
\newblock \bibinfo{journal}{\emph{Journal of Broadcasting \& Electronic Media}}
  \bibinfo{volume}{54}, \bibinfo{number}{1} (\bibinfo{year}{2010}),
  \bibinfo{pages}{40--53}.
\newblock


\bibitem[\protect\citeauthoryear{Entman}{Entman}{1992}]%
        {entman1992blacks}
\bibfield{author}{\bibinfo{person}{Robert~M. Entman}.}
  \bibinfo{year}{1992}\natexlab{}.
\newblock \showarticletitle{Blacks in the news: Television, modern racism and
  cultural change}.
\newblock \bibinfo{journal}{\emph{Journalism Quarterly}} \bibinfo{volume}{69},
  \bibinfo{number}{2} (\bibinfo{year}{1992}), \bibinfo{pages}{341--361}.
\newblock


\bibitem[\protect\citeauthoryear{Face}{Face}{2021}]%
        {neuralcoref}
\bibfield{author}{\bibinfo{person}{Hugging Face}.}
  \bibinfo{year}{2021}\natexlab{}.
\newblock \bibinfo{title}{NeuralCoref 4.0: Coreference Resolution in spaCy with
  Neural Networks}.
\newblock
\newblock
\urldef\tempurl%
\url{https://github.com/huggingface/neuralcoref}
\showURL{%
\tempurl}
\newblock
\shownote{Accessed: 2022-02-04.}


\bibitem[\protect\citeauthoryear{Fast, Vachovsky, and Bernstein}{Fast
  et~al\mbox{.}}{2016}]%
        {fast2016shirtless}
\bibfield{author}{\bibinfo{person}{Ethan Fast}, \bibinfo{person}{Tina
  Vachovsky}, {and} \bibinfo{person}{Michael~S. Bernstein}.}
  \bibinfo{year}{2016}\natexlab{}.
\newblock \showarticletitle{Shirtless and dangerous: Quantifying linguistic
  signals of gender bias in an online fiction writing community}. In
  \bibinfo{booktitle}{\emph{Proceedings of the International AAAI Conference on
  Web and Social Media}}. \bibinfo{publisher}{Association for Computational
  Linguistics}, \bibinfo{address}{Florence, Italy}, \bibinfo{pages}{112--120}.
\newblock


\bibitem[\protect\citeauthoryear{Ferrer, van Nuenen, Such, and Criado}{Ferrer
  et~al\mbox{.}}{2021}]%
        {ferrer2020discovering}
\bibfield{author}{\bibinfo{person}{Xavier Ferrer}, \bibinfo{person}{Tom van
  Nuenen}, \bibinfo{person}{Jose~M Such}, {and} \bibinfo{person}{Natalia
  Criado}.} \bibinfo{year}{2021}\natexlab{}.
\newblock \showarticletitle{Discovering and categorising language biases in
  Reddit}.
\newblock \bibinfo{journal}{\emph{Proceedings of the International AAAI
  Conference on Web and Social Media}}, \bibinfo{pages}{140--151}.
\newblock


\bibitem[\protect\citeauthoryear{Flerx, Fidler, and Rogers}{Flerx
  et~al\mbox{.}}{1976}]%
        {flerx1976sex}
\bibfield{author}{\bibinfo{person}{Vicki~C Flerx}, \bibinfo{person}{Dorothy~S
  Fidler}, {and} \bibinfo{person}{Ronald~W Rogers}.}
  \bibinfo{year}{1976}\natexlab{}.
\newblock \showarticletitle{Sex role stereotypes: Developmental aspects and
  early intervention}.
\newblock \bibinfo{journal}{\emph{Child Development}} (\bibinfo{year}{1976}),
  \bibinfo{pages}{998--1007}.
\newblock


\bibitem[\protect\citeauthoryear{for Literacy~in Primary~Education}{for
  Literacy~in Primary~Education}{2020}]%
        {CLPE}
\bibfield{author}{\bibinfo{person}{The~Centre for Literacy~in
  Primary~Education}.} \bibinfo{year}{2020}\natexlab{}.
\newblock \bibinfo{title}{Reflecting Realities-Survey of Ethnic Representation
  within UK Children’s Literature}.
\newblock
\newblock
\urldef\tempurl%
\url{https://clpe.org.uk/research/clpe-reflecting-realities-survey-ethnic-representation-within-uk-childrens-literature}
\showURL{%
\tempurl}
\newblock
\shownote{Accessed: 08/30/2021.}


\bibitem[\protect\citeauthoryear{Gad, Javed, Ghani, Elmqvist, Ewing, Hampton,
  and Ramakrishnan}{Gad et~al\mbox{.}}{2015}]%
        {DBLP:journals/tvcg/GadJGEEHR15}
\bibfield{author}{\bibinfo{person}{Samah Gad}, \bibinfo{person}{Waqas Javed},
  \bibinfo{person}{Sohaib Ghani}, \bibinfo{person}{Niklas Elmqvist},
  \bibinfo{person}{E.~Thomas Ewing}, \bibinfo{person}{Keith~N. Hampton}, {and}
  \bibinfo{person}{Naren Ramakrishnan}.} \bibinfo{year}{2015}\natexlab{}.
\newblock \showarticletitle{{ThemeDelta}: Dynamic Segmentations over Temporal
  Topic Models}.
\newblock \bibinfo{journal}{\emph{{{IEEE} Transactions on Visualization and
  Computer Graphics}}} \bibinfo{volume}{21}, \bibinfo{number}{5}
  (\bibinfo{year}{2015}), \bibinfo{pages}{672--685}.
\newblock
\urldef\tempurl%
\url{https://doi.org/10.1109/TVCG.2014.2388208}
\showDOI{\tempurl}


\bibitem[\protect\citeauthoryear{Gan, Zhu, Li, Liang, Cao, and Zhou}{Gan
  et~al\mbox{.}}{2014}]%
        {Qihong2014}
\bibfield{author}{\bibinfo{person}{Qihong Gan}, \bibinfo{person}{Min Zhu},
  \bibinfo{person}{Mingzhao Li}, \bibinfo{person}{Ting Liang},
  \bibinfo{person}{Yu Cao}, {and} \bibinfo{person}{Baoyao Zhou}.}
  \bibinfo{year}{2014}\natexlab{}.
\newblock \showarticletitle{Document visualization: an overview of current
  research}.
\newblock \bibinfo{journal}{\emph{WIREs Computational Statistics}}
  \bibinfo{volume}{6}, \bibinfo{number}{1} (\bibinfo{year}{2014}),
  \bibinfo{pages}{19--36}.
\newblock
\urldef\tempurl%
\url{https://doi.org/10.1002/wics.1285}
\showDOI{\tempurl}


\bibitem[\protect\citeauthoryear{Garg, Schiebinger, Jurafsky, and Zou}{Garg
  et~al\mbox{.}}{2018}]%
        {100years}
\bibfield{author}{\bibinfo{person}{Nikhil Garg}, \bibinfo{person}{Londa
  Schiebinger}, \bibinfo{person}{Dan Jurafsky}, {and} \bibinfo{person}{James
  Zou}.} \bibinfo{year}{2018}\natexlab{}.
\newblock \showarticletitle{Word embeddings quantify 100 years of gender and
  ethnic stereotypes}.
\newblock \bibinfo{journal}{\emph{Proceedings of the National Academy of
  Sciences}} \bibinfo{volume}{115}, \bibinfo{number}{16}
  (\bibinfo{year}{2018}), \bibinfo{pages}{E3635--E3644}.
\newblock


\bibitem[\protect\citeauthoryear{Garimella, Banea, Hovy, and
  Mihalcea}{Garimella et~al\mbox{.}}{2019}]%
        {garimella2019women}
\bibfield{author}{\bibinfo{person}{Aparna Garimella}, \bibinfo{person}{Carmen
  Banea}, \bibinfo{person}{Dirk Hovy}, {and} \bibinfo{person}{Rada Mihalcea}.}
  \bibinfo{year}{2019}\natexlab{}.
\newblock \showarticletitle{Women’s syntactic resilience and men’s
  grammatical luck: Gender-Bias in Part-of-Speech Tagging and Dependency
  Parsing}. In \bibinfo{booktitle}{\emph{Proceedings of the Annual Meeting of
  the Association for Computational Linguistics}}.
  \bibinfo{publisher}{Association for Computational Linguistics},
  \bibinfo{address}{Florence, Italy}, \bibinfo{pages}{3493--3498}.
\newblock


\bibitem[\protect\citeauthoryear{Ghai, Hoque, and Mueller}{Ghai
  et~al\mbox{.}}{2021}]%
        {ghai2021wordbias}
\bibfield{author}{\bibinfo{person}{Bhavya Ghai}, \bibinfo{person}{Md~Naimul
  Hoque}, {and} \bibinfo{person}{Klaus Mueller}.}
  \bibinfo{year}{2021}\natexlab{}.
\newblock \showarticletitle{WordBias: An Interactive Visual Tool for
  Discovering Intersectional Biases Encoded in Word Embeddings}. In
  \bibinfo{booktitle}{\emph{Extended Abstracts of the ACM Conference on Human
  Factors in Computing Systems}}. \bibinfo{publisher}{{ACM}},
  \bibinfo{address}{New York, NY, USA}, \bibinfo{pages}{1--7}.
\newblock


\bibitem[\protect\citeauthoryear{Ghai, Mishra, and Mueller}{Ghai
  et~al\mbox{.}}{2022}]%
        {ghai2022cascaded}
\bibfield{author}{\bibinfo{person}{Bhavya Ghai}, \bibinfo{person}{Mihir
  Mishra}, {and} \bibinfo{person}{Klaus Mueller}.}
  \bibinfo{year}{2022}\natexlab{}.
\newblock \bibinfo{title}{Cascaded Debiasing : Studying the Cumulative Effect
  of Multiple Fairness-Enhancing Interventions}.
\newblock , \bibinfo{numpages}{12}~pages.
\newblock
\showeprint[arXiv]{2202.03734}
\urldef\tempurl%
\url{https://arxiv.org/abs/2202.03734}
\showURL{%
\tempurl}


\bibitem[\protect\citeauthoryear{Granthika}{Granthika}{2021}]%
        {granthika}
\bibfield{author}{\bibinfo{person}{Granthika}.}
  \bibinfo{year}{2021}\natexlab{}.
\newblock \bibinfo{title}{Granthika Writing Tool}.
\newblock
\newblock
\urldef\tempurl%
\url{https://granthika.co}
\showURL{%
\tempurl}
\newblock
\shownote{Accessed: 2022-02-04.}


\bibitem[\protect\citeauthoryear{Gronemann, J{\"{u}}nger, Liers, and
  Mambelli}{Gronemann et~al\mbox{.}}{2016}]%
        {DBLP:conf/gd/GronemannJLM16}
\bibfield{author}{\bibinfo{person}{Martin Gronemann}, \bibinfo{person}{Michael
  J{\"{u}}nger}, \bibinfo{person}{Frauke Liers}, {and}
  \bibinfo{person}{Francesco Mambelli}.} \bibinfo{year}{2016}\natexlab{}.
\newblock \showarticletitle{Crossing Minimization in Storyline Visualization}.
  In \bibinfo{booktitle}{\emph{Proceedings of the International Symposium on
  Graph Drawing and Network Visualization}} \emph{(\bibinfo{series}{Lecture
  Notes in Computer Science}, Vol.~\bibinfo{volume}{9801})}.
  \bibinfo{publisher}{Springer}, \bibinfo{pages}{367--381}.
\newblock
\urldef\tempurl%
\url{https://doi.org/10.1007/978-3-319-50106-2\_29}
\showDOI{\tempurl}


\bibitem[\protect\citeauthoryear{Hamilton, Anderson, Broaddus, and
  Young}{Hamilton et~al\mbox{.}}{2006}]%
        {hamilton2006gender}
\bibfield{author}{\bibinfo{person}{Mykol~C. Hamilton}, \bibinfo{person}{David
  Anderson}, \bibinfo{person}{Michelle Broaddus}, {and} \bibinfo{person}{Kate
  Young}.} \bibinfo{year}{2006}\natexlab{}.
\newblock \showarticletitle{Gender stereotyping and under-representation of
  female characters in 200 popular children’s picture books: A twenty-first
  century update}.
\newblock \bibinfo{journal}{\emph{Sex Roles}} \bibinfo{volume}{55},
  \bibinfo{number}{11} (\bibinfo{year}{2006}), \bibinfo{pages}{757--765}.
\newblock


\bibitem[\protect\citeauthoryear{Hart}{Hart}{2006}]%
        {hart2006nasa}
\bibfield{author}{\bibinfo{person}{Sandra~G Hart}.}
  \bibinfo{year}{2006}\natexlab{}.
\newblock \showarticletitle{NASA-task load index (NASA-TLX); 20 years later}.
\newblock \bibinfo{journal}{\emph{Proceedings of the human factors and
  ergonomics society annual meeting}} \bibinfo{volume}{50}, \bibinfo{number}{9}
  (\bibinfo{year}{2006}), \bibinfo{pages}{904--908}.
\newblock


\bibitem[\protect\citeauthoryear{Hearst, Pedersen, Patil, Lee, Laskowski, and
  Franconeri}{Hearst et~al\mbox{.}}{2019}]%
        {hearst2019evaluation}
\bibfield{author}{\bibinfo{person}{Marti~A. Hearst}, \bibinfo{person}{Emily
  Pedersen}, \bibinfo{person}{Lekha Patil}, \bibinfo{person}{Elsie Lee},
  \bibinfo{person}{Paul Laskowski}, {and} \bibinfo{person}{Steven Franconeri}.}
  \bibinfo{year}{2019}\natexlab{}.
\newblock \showarticletitle{An evaluation of semantically grouped word cloud
  designs}.
\newblock \bibinfo{journal}{\emph{{{IEEE} Transactions on Visualization and
  Computer Graphics}}} \bibinfo{volume}{26}, \bibinfo{number}{9}
  (\bibinfo{year}{2019}), \bibinfo{pages}{2748--2761}.
\newblock


\bibitem[\protect\citeauthoryear{Honnibal, Montani, Van~Landeghem, and
  Boyd}{Honnibal et~al\mbox{.}}{2020}]%
        {spacy}
\bibfield{author}{\bibinfo{person}{Matthew Honnibal}, \bibinfo{person}{Ines
  Montani}, \bibinfo{person}{Sofie Van~Landeghem}, {and}
  \bibinfo{person}{Adriane Boyd}.} \bibinfo{year}{2020}\natexlab{}.
\newblock \bibinfo{booktitle}{\emph{{spaCy: Industrial-strength Natural
  Language Processing in Python}}}.
\newblock
\urldef\tempurl%
\url{https://doi.org/10.5281/zenodo.1212303}
\showDOI{\tempurl}


\bibitem[\protect\citeauthoryear{Hoque, Saquib, Billah, and Mueller}{Hoque
  et~al\mbox{.}}{2020}]%
        {hoque2020toward}
\bibfield{author}{\bibinfo{person}{Md~Naimul Hoque}, \bibinfo{person}{Nazmus
  Saquib}, \bibinfo{person}{Syed~Masum Billah}, {and} \bibinfo{person}{Klaus
  Mueller}.} \bibinfo{year}{2020}\natexlab{}.
\newblock \showarticletitle{Toward Interactively Balancing the Screen Time of
  Actors Based on Observable Phenotypic Traits in Live Telecast}.
\newblock \bibinfo{journal}{\emph{Proceedings of the ACM on Human-Computer
  Interaction}} \bibinfo{volume}{4}, \bibinfo{number}{CSCW2}
  (\bibinfo{year}{2020}), \bibinfo{pages}{1--18}.
\newblock


\bibitem[\protect\citeauthoryear{Hoyle, Wolf-Sonkin, Wallach, Augenstein, and
  Cotterell}{Hoyle et~al\mbox{.}}{2019}]%
        {hoyle-etal-2019-unsupervised}
\bibfield{author}{\bibinfo{person}{Alexander~Miserlis Hoyle},
  \bibinfo{person}{Lawrence Wolf-Sonkin}, \bibinfo{person}{Hanna Wallach},
  \bibinfo{person}{Isabelle Augenstein}, {and} \bibinfo{person}{Ryan
  Cotterell}.} \bibinfo{year}{2019}\natexlab{}.
\newblock \showarticletitle{Unsupervised Discovery of Gendered Language through
  Latent-Variable Modeling}. In \bibinfo{booktitle}{\emph{Proceedings of the
  Annual Meeting of the Association for Computational Linguistics}}.
  \bibinfo{publisher}{Association for Computational Linguistics},
  \bibinfo{address}{Florence, Italy}, \bibinfo{pages}{1706--1716}.
\newblock
\urldef\tempurl%
\url{https://doi.org/10.18653/v1/P19-1167}
\showDOI{\tempurl}


\bibitem[\protect\citeauthoryear{Huang, Li, and Shan}{Huang
  et~al\mbox{.}}{2013}]%
        {DBLP:conf/taai/HuangLS13}
\bibfield{author}{\bibinfo{person}{Chieh{-}Jen Huang},
  \bibinfo{person}{Cheng{-}Te Li}, {and} \bibinfo{person}{Man{-}Kwan Shan}.}
  \bibinfo{year}{2013}\natexlab{}.
\newblock \showarticletitle{{VizStory}: Visualization of Digital Narrative for
  Fairy Tales}. In \bibinfo{booktitle}{\emph{Proceedings of the Conference on
  Technologies and Applications of Artificial Intelligence}}.
  \bibinfo{publisher}{{IEEE} Computer Society}, \bibinfo{address}{Piscataway,
  NJ, USA}, \bibinfo{pages}{67--72}.
\newblock
\urldef\tempurl%
\url{https://doi.org/10.1109/TAAI.2013.26}
\showDOI{\tempurl}


\bibitem[\protect\citeauthoryear{Hui, Gergle, and Gerber}{Hui
  et~al\mbox{.}}{2018}]%
        {hui2018introassist}
\bibfield{author}{\bibinfo{person}{Julie~S Hui}, \bibinfo{person}{Darren
  Gergle}, {and} \bibinfo{person}{Elizabeth~M Gerber}.}
  \bibinfo{year}{2018}\natexlab{}.
\newblock \showarticletitle{IntroAssist: A tool to support writing introductory
  help requests}. In \bibinfo{booktitle}{\emph{Proceedings of the ACM
  Conference on Human Factors in Computing Systems}}.
  \bibinfo{publisher}{{ACM}}, \bibinfo{address}{New York, NY, USA},
  \bibinfo{pages}{1--13}.
\newblock
\urldef\tempurl%
\url{https://doi.org/10.1145/3173574.3173596}
\showDOI{\tempurl}


\bibitem[\protect\citeauthoryear{Hunt and Ramón}{Hunt and Ramón}{2020}]%
        {uclareport2020}
\bibfield{author}{\bibinfo{person}{Darnell Hunt} {and}
  \bibinfo{person}{Ana-Christina Ramón}.} \bibinfo{year}{2020}\natexlab{}.
\newblock \bibinfo{title}{Hollywood Diversity Report 2020}.
\newblock
\newblock
\urldef\tempurl%
\url{https://socialsciences.ucla.edu/wp-content/uploads/2020/02/UCLA-Hollywood-Diversity-Report-2020-Film-2-6-2020.pdf}
\showURL{%
\tempurl}


\bibitem[\protect\citeauthoryear{Joseph, Wei, and Carley}{Joseph
  et~al\mbox{.}}{2017}]%
        {joseph2017girls}
\bibfield{author}{\bibinfo{person}{Kenneth Joseph}, \bibinfo{person}{Wei Wei},
  {and} \bibinfo{person}{Kathleen~M. Carley}.} \bibinfo{year}{2017}\natexlab{}.
\newblock \showarticletitle{Girls rule, boys drool: Extracting semantic and
  affective stereotypes from Twitter}. In \bibinfo{booktitle}{\emph{Proceedings
  of the ACM Conference on Computer Supported Cooperative Work and Social
  Computing}}. \bibinfo{publisher}{{ACM}}, \bibinfo{address}{New York, NY,
  USA}, \bibinfo{pages}{1362--1374}.
\newblock


\bibitem[\protect\citeauthoryear{Jurafsky and Martin}{Jurafsky and
  Martin}{[n.d.]}]%
        {jurafskyspeech}
\bibfield{author}{\bibinfo{person}{Daniel Jurafsky} {and}
  \bibinfo{person}{James~H Martin}.} \bibinfo{year}{[n.d.]}\natexlab{}.
\newblock \bibinfo{title}{Speech and Language Processing: An Introduction to
  Natural Language Processing, Computational Linguistics, and Speech
  Recognition}.
\newblock
\newblock


\bibitem[\protect\citeauthoryear{Keim and Oelke}{Keim and Oelke}{2007}]%
        {DBLP:conf/ieeevast/KeimO07}
\bibfield{author}{\bibinfo{person}{Daniel~A. Keim} {and}
  \bibinfo{person}{Daniela Oelke}.} \bibinfo{year}{2007}\natexlab{}.
\newblock \showarticletitle{Literature Fingerprinting: {A} New Method for
  Visual Literary Analysis}. In \bibinfo{booktitle}{\emph{Proceedings of the
  {IEEE} Symposium on Visual Analytics Science and Technology}}.
  \bibinfo{publisher}{{IEEE}}, \bibinfo{address}{Piscataway, NJ, USA},
  \bibinfo{pages}{115--122}.
\newblock
\urldef\tempurl%
\url{https://doi.org/10.1109/VAST.2007.4389004}
\showDOI{\tempurl}


\bibitem[\protect\citeauthoryear{Kim, Pethe, and Skiena}{Kim
  et~al\mbox{.}}{2020}]%
        {kim2020time}
\bibfield{author}{\bibinfo{person}{Allen Kim}, \bibinfo{person}{Charuta Pethe},
  {and} \bibinfo{person}{Steven Skiena}.} \bibinfo{year}{2020}\natexlab{}.
\newblock \showarticletitle{What Time Is It? Temporal Analysis of Novels}. In
  \bibinfo{booktitle}{\emph{Proceedings of the Conference on Empirical Methods
  in Natural Language Processing}}. \bibinfo{publisher}{Association for
  Computational Linguistics}, \bibinfo{address}{Florence, Italy},
  \bibinfo{pages}{9076--9086}.
\newblock


\bibitem[\protect\citeauthoryear{Kim, Bach, Im, Schriber, Gross, and
  Pfister}{Kim et~al\mbox{.}}{2018}]%
        {DBLP:journals/tvcg/KimBISGP18}
\bibfield{author}{\bibinfo{person}{Nam~Wook Kim}, \bibinfo{person}{Benjamin
  Bach}, \bibinfo{person}{Hyejin Im}, \bibinfo{person}{Sasha Schriber},
  \bibinfo{person}{Markus~H. Gross}, {and} \bibinfo{person}{Hanspeter
  Pfister}.} \bibinfo{year}{2018}\natexlab{}.
\newblock \showarticletitle{Visualizing Nonlinear Narratives with Story
  Curves}.
\newblock \bibinfo{journal}{\emph{{{IEEE} Transactions on Visualization and
  Computer Graphics}}} \bibinfo{volume}{24}, \bibinfo{number}{1}
  (\bibinfo{year}{2018}), \bibinfo{pages}{595--604}.
\newblock
\urldef\tempurl%
\url{https://doi.org/10.1109/TVCG.2017.2744118}
\showDOI{\tempurl}


\bibitem[\protect\citeauthoryear{Kolbe and La~Voie}{Kolbe and La~Voie}{1981}]%
        {kolbe1981sex}
\bibfield{author}{\bibinfo{person}{Richard Kolbe} {and}
  \bibinfo{person}{Joseph~C La~Voie}.} \bibinfo{year}{1981}\natexlab{}.
\newblock \showarticletitle{Sex-role stereotyping in preschool children's
  picture books}.
\newblock \bibinfo{journal}{\emph{Social Psychology Quarterly}}
  (\bibinfo{year}{1981}), \bibinfo{pages}{369--374}.
\newblock


\bibitem[\protect\citeauthoryear{Kraicer and Piper}{Kraicer and Piper}{2019}]%
        {kraicer2019social}
\bibfield{author}{\bibinfo{person}{Eve Kraicer} {and} \bibinfo{person}{Andrew
  Piper}.} \bibinfo{year}{2019}\natexlab{}.
\newblock \showarticletitle{Social characters: the hierarchy of gender in
  contemporary English-language fiction}.
\newblock \bibinfo{journal}{\emph{Journal of Cultural Analytics}}
  \bibinfo{volume}{1}, \bibinfo{number}{1} (\bibinfo{year}{2019}),
  \bibinfo{pages}{11055}.
\newblock


\bibitem[\protect\citeauthoryear{Labatut and Bost}{Labatut and Bost}{2019}]%
        {labatut2019extraction}
\bibfield{author}{\bibinfo{person}{Vincent Labatut} {and}
  \bibinfo{person}{Xavier Bost}.} \bibinfo{year}{2019}\natexlab{}.
\newblock \showarticletitle{Extraction and analysis of fictional character
  networks: A survey}.
\newblock \bibinfo{journal}{\emph{Comput. Surveys}} \bibinfo{volume}{52},
  \bibinfo{number}{5} (\bibinfo{year}{2019}), \bibinfo{pages}{1--40}.
\newblock


\bibitem[\protect\citeauthoryear{Layne and Alemanji}{Layne and
  Alemanji}{2015}]%
        {layne2015zebra}
\bibfield{author}{\bibinfo{person}{Heidi Layne} {and}
  \bibinfo{person}{Amikeng~A Alemanji}.} \bibinfo{year}{2015}\natexlab{}.
\newblock \showarticletitle{``Zebra world'': The promotion of imperial
  stereotypes in a children’s book}.
\newblock \bibinfo{journal}{\emph{Power and Education}} \bibinfo{volume}{7},
  \bibinfo{number}{2} (\bibinfo{year}{2015}), \bibinfo{pages}{181--195}.
\newblock


\bibitem[\protect\citeauthoryear{Lee, Liang, and Yang}{Lee
  et~al\mbox{.}}{2022}]%
        {lee2022coauthor}
\bibfield{author}{\bibinfo{person}{Mina Lee}, \bibinfo{person}{Percy Liang},
  {and} \bibinfo{person}{Qian Yang}.} \bibinfo{year}{2022}\natexlab{}.
\newblock \showarticletitle{CoAuthor: Designing a Human-AI Collaborative
  Writing Dataset for Exploring Language Model Capabilities}.
\newblock \bibinfo{journal}{\emph{procCHI}} (\bibinfo{year}{2022}).
\newblock


\bibitem[\protect\citeauthoryear{Literature and Latte}{Literature and
  Latte}{2021}]%
        {scrivener}
\bibfield{author}{\bibinfo{person}{Literature} {and} \bibinfo{person}{Latte}.}
  \bibinfo{year}{2021}\natexlab{}.
\newblock \bibinfo{title}{Scrivener}.
\newblock
\newblock
\urldef\tempurl%
\url{https://www.literatureandlatte.com/scrivener/overview/}
\showURL{%
\tempurl}
\newblock
\shownote{Accessed: 2022-02-04.}


\bibitem[\protect\citeauthoryear{Liu, Wu, Wei, Liu, and Liu}{Liu
  et~al\mbox{.}}{2013}]%
        {DBLP:journals/tvcg/LiuWWLL13}
\bibfield{author}{\bibinfo{person}{Shixia Liu}, \bibinfo{person}{Yingcai Wu},
  \bibinfo{person}{Enxun Wei}, \bibinfo{person}{Mengchen Liu}, {and}
  \bibinfo{person}{Yang Liu}.} \bibinfo{year}{2013}\natexlab{}.
\newblock \showarticletitle{{StoryFlow}: Tracking the Evolution of Stories}.
\newblock \bibinfo{journal}{\emph{{{IEEE} Transactions on Visualization and
  Computer Graphics}}} \bibinfo{volume}{19}, \bibinfo{number}{12}
  (\bibinfo{year}{2013}), \bibinfo{pages}{2436--2445}.
\newblock
\urldef\tempurl%
\url{https://doi.org/10.1109/TVCG.2013.196}
\showDOI{\tempurl}


\bibitem[\protect\citeauthoryear{Lucy, Demszky, Bromley, and Jurafsky}{Lucy
  et~al\mbox{.}}{2020}]%
        {lucy2020content}
\bibfield{author}{\bibinfo{person}{Li Lucy}, \bibinfo{person}{Dorottya
  Demszky}, \bibinfo{person}{Patricia Bromley}, {and} \bibinfo{person}{Dan
  Jurafsky}.} \bibinfo{year}{2020}\natexlab{}.
\newblock \showarticletitle{Content analysis of textbooks via natural language
  processing: Findings on gender, race, and ethnicity in Texas US history
  textbooks}.
\newblock \bibinfo{journal}{\emph{AERA Open}} \bibinfo{volume}{6},
  \bibinfo{number}{3} (\bibinfo{year}{2020}),
  \bibinfo{pages}{2332858420940312}.
\newblock


\bibitem[\protect\citeauthoryear{Madaan, Mehta, Agrawaal, Malhotra, Aggarwal,
  Gupta, and Saxena}{Madaan et~al\mbox{.}}{2018}]%
        {madaan2018analyze}
\bibfield{author}{\bibinfo{person}{Nishtha Madaan}, \bibinfo{person}{Sameep
  Mehta}, \bibinfo{person}{Taneea Agrawaal}, \bibinfo{person}{Vrinda Malhotra},
  \bibinfo{person}{Aditi Aggarwal}, \bibinfo{person}{Yatin Gupta}, {and}
  \bibinfo{person}{Mayank Saxena}.} \bibinfo{year}{2018}\natexlab{}.
\newblock \showarticletitle{Analyze, detect and remove gender stereotyping from
  bollywood movies}. In \bibinfo{booktitle}{\emph{Conference on fairness,
  accountability and transparency}}. PMLR, \bibinfo{pages}{92--105}.
\newblock


\bibitem[\protect\citeauthoryear{Maiden, Zachos, Brown, Brock, Nyre,
  Nyg{\aa}rd~Tonheim, Apsotolou, and Evans}{Maiden et~al\mbox{.}}{2018}]%
        {maiden2018making}
\bibfield{author}{\bibinfo{person}{Neil Maiden}, \bibinfo{person}{Konstantinos
  Zachos}, \bibinfo{person}{Amanda Brown}, \bibinfo{person}{George Brock},
  \bibinfo{person}{Lars Nyre}, \bibinfo{person}{Aleksander Nyg{\aa}rd~Tonheim},
  \bibinfo{person}{Dimitris Apsotolou}, {and} \bibinfo{person}{Jeremy Evans}.}
  \bibinfo{year}{2018}\natexlab{}.
\newblock \showarticletitle{Making the news: Digital creativity support for
  journalists}. In \bibinfo{booktitle}{\emph{Proceedings of the {ACM}
  Conference on Human Factors in Computing Systems}}.
  \bibinfo{publisher}{{ACM}}, \bibinfo{address}{New York, NY, USA},
  \bibinfo{pages}{1--11}.
\newblock


\bibitem[\protect\citeauthoryear{Matamoros-Fern{\'a}ndez}{Matamoros-Fern{\'a}ndez}{2017}]%
        {matamoros2017platformed}
\bibfield{author}{\bibinfo{person}{Ariadna Matamoros-Fern{\'a}ndez}.}
  \bibinfo{year}{2017}\natexlab{}.
\newblock \showarticletitle{Platformed racism: The mediation and circulation of
  an Australian race-based controversy on Twitter, Facebook and YouTube}.
\newblock \bibinfo{journal}{\emph{Information, Communication \& Society}}
  \bibinfo{volume}{20}, \bibinfo{number}{6} (\bibinfo{year}{2017}),
  \bibinfo{pages}{930--946}.
\newblock


\bibitem[\protect\citeauthoryear{McCurdy, Lein, Coles, and Meyer}{McCurdy
  et~al\mbox{.}}{2016}]%
        {DBLP:journals/tvcg/McCurdyLCM16}
\bibfield{author}{\bibinfo{person}{Nina McCurdy}, \bibinfo{person}{Julie Lein},
  \bibinfo{person}{Katherine Coles}, {and} \bibinfo{person}{Miriah~D. Meyer}.}
  \bibinfo{year}{2016}\natexlab{}.
\newblock \showarticletitle{Poemage: Visualizing the Sonic Topology of a Poem}.
\newblock \bibinfo{journal}{\emph{{{IEEE} Transactions on Visualization and
  Computer Graphics}}} \bibinfo{volume}{22}, \bibinfo{number}{1}
  (\bibinfo{year}{2016}), \bibinfo{pages}{439--448}.
\newblock
\urldef\tempurl%
\url{https://doi.org/10.1109/TVCG.2015.2467811}
\showDOI{\tempurl}


\bibitem[\protect\citeauthoryear{Mehrabi, Gowda, Morstatter, Peng, and
  Galstyan}{Mehrabi et~al\mbox{.}}{2020}]%
        {mehrabi2020man}
\bibfield{author}{\bibinfo{person}{Ninareh Mehrabi}, \bibinfo{person}{Thamme
  Gowda}, \bibinfo{person}{Fred Morstatter}, \bibinfo{person}{Nanyun Peng},
  {and} \bibinfo{person}{Aram Galstyan}.} \bibinfo{year}{2020}\natexlab{}.
\newblock \showarticletitle{Man is to person as woman is to location: Measuring
  gender bias in named entity recognition}. In
  \bibinfo{booktitle}{\emph{Proceedings of the ACM Conference on Hypertext and
  Social Media}}. \bibinfo{publisher}{{ACM}}, \bibinfo{address}{New York, NY,
  USA}, \bibinfo{pages}{231--232}.
\newblock


\bibitem[\protect\citeauthoryear{Mishra, He, and Belli}{Mishra
  et~al\mbox{.}}{2020}]%
        {mishra2020assessing}
\bibfield{author}{\bibinfo{person}{Shubhanshu Mishra}, \bibinfo{person}{Sijun
  He}, {and} \bibinfo{person}{Luca Belli}.} \bibinfo{year}{2020}\natexlab{}.
\newblock \showarticletitle{Assessing Demographic Bias in Named Entity
  Recognition}.
\newblock \bibinfo{journal}{\emph{arXiv preprint arXiv:2008.03415}}
  (\bibinfo{year}{2020}).
\newblock


\bibitem[\protect\citeauthoryear{Mitri}{Mitri}{2020}]%
        {mitri2020story}
\bibfield{author}{\bibinfo{person}{Michel Mitri}.}
  \bibinfo{year}{2020}\natexlab{}.
\newblock \showarticletitle{Story analysis using natural language processing
  and interactive dashboards}.
\newblock \bibinfo{journal}{\emph{Journal of Computer Information Systems}}
  (\bibinfo{year}{2020}), \bibinfo{pages}{1--11}.
\newblock


\bibitem[\protect\citeauthoryear{Narahara}{Narahara}{1998}]%
        {narahara1998gender}
\bibfield{author}{\bibinfo{person}{May~M Narahara}.}
  \bibinfo{year}{1998}\natexlab{}.
\newblock \showarticletitle{Gender Stereotypes in Children's Picture Books.}
\newblock  (\bibinfo{year}{1998}).
\newblock


\bibitem[\protect\citeauthoryear{Norberg}{Norberg}{2016}]%
        {norberg2016naughty}
\bibfield{author}{\bibinfo{person}{Cathrine Norberg}.}
  \bibinfo{year}{2016}\natexlab{}.
\newblock \showarticletitle{Naughty boys and sexy girls: the representation of
  young individuals in a web-based corpus of English}.
\newblock \bibinfo{journal}{\emph{Journal of English Linguistics}}
  \bibinfo{volume}{44}, \bibinfo{number}{4} (\bibinfo{year}{2016}),
  \bibinfo{pages}{291--317}.
\newblock


\bibitem[\protect\citeauthoryear{on~Gender~in Media}{on~Gender~in
  Media}{2008}]%
        {gdfilms2008}
\bibfield{author}{\bibinfo{person}{Geena Davis~Institue on~Gender~in Media}.}
  \bibinfo{year}{2008}\natexlab{}.
\newblock \bibinfo{title}{Gender Stereotypes: An Analysis of Popular Films and
  TV}.
\newblock
\newblock
\urldef\tempurl%
\url{https://seejane.org/wp-content/uploads/GDIGM_Gender_Stereotypes.pdf}
\showURL{%
\tempurl}


\bibitem[\protect\citeauthoryear{on~Gender~in Media}{on~Gender~in
  Media}{2019a}]%
        {parity_children}
\bibfield{author}{\bibinfo{person}{Geena Davis~Institue on~Gender~in Media}.}
  \bibinfo{year}{2019}\natexlab{a}.
\newblock \bibinfo{title}{Historic Gender Parity in Children's Television}.
\newblock
\newblock
\urldef\tempurl%
\url{https://seejane.org/research-informs-empowers/see-jane-2019/}
\showURL{%
\tempurl}


\bibitem[\protect\citeauthoryear{on~Gender~in Media}{on~Gender~in
  Media}{2019b}]%
        {gdblack2019}
\bibfield{author}{\bibinfo{person}{Geena Davis~Institue on~Gender~in Media}.}
  \bibinfo{year}{2019}\natexlab{b}.
\newblock \bibinfo{title}{Representation Of Black Women in Hollywood}.
\newblock
\newblock
\urldef\tempurl%
\url{https://seejane.org/research-informs-empowers/representations-of-black-women-in-hollywood/}
\showURL{%
\tempurl}


\bibitem[\protect\citeauthoryear{on~Gender~in Media}{on~Gender~in
  Media}{2020}]%
        {parity_films}
\bibfield{author}{\bibinfo{person}{Geena Davis~Institue on~Gender~in Media}.}
  \bibinfo{year}{2020}\natexlab{}.
\newblock \bibinfo{title}{Historic Gender Parity in Family Films!}
\newblock
\newblock
\urldef\tempurl%
\url{https://seejane.org/2020-film-historic-gender-parity-in-family-films/}
\showURL{%
\tempurl}


\bibitem[\protect\citeauthoryear{on~Gender~in Media}{on~Gender~in
  Media}{2021}]%
        {gd_equity2021}
\bibfield{author}{\bibinfo{person}{Geena Davis~Institue on~Gender~in Media}.}
  \bibinfo{year}{2021}\natexlab{}.
\newblock \bibinfo{title}{Behind the Scenes: the State of Inclusion and Equity
  In TV Writing}.
\newblock
\newblock
\urldef\tempurl%
\url{https://seejane.org/research-informs-empowers/behind-the-scenes-the-state-of-inclusion-and-equity-in-tv-writing/}
\showURL{%
\tempurl}


\bibitem[\protect\citeauthoryear{Paynter}{Paynter}{2011}]%
        {paynter2011gender}
\bibfield{author}{\bibinfo{person}{Kelly~Crisp Paynter}.}
  \bibinfo{year}{2011}\natexlab{}.
\newblock \bibinfo{booktitle}{\emph{Gender stereotypes and representation of
  female characters in children's picture books}}.
\newblock \bibinfo{publisher}{Liberty University}.
\newblock


\bibitem[\protect\citeauthoryear{Pearce}{Pearce}{2008}]%
        {pearce2008investigating}
\bibfield{author}{\bibinfo{person}{Michael Pearce}.}
  \bibinfo{year}{2008}\natexlab{}.
\newblock \showarticletitle{Investigating the collocational behaviour of MAN
  and WOMAN in the BNC using Sketch Engine}.
\newblock \bibinfo{journal}{\emph{Corpora}} \bibinfo{volume}{3},
  \bibinfo{number}{1} (\bibinfo{year}{2008}), \bibinfo{pages}{1--29}.
\newblock


\bibitem[\protect\citeauthoryear{Peterson and Lach}{Peterson and Lach}{1990}]%
        {peterson1990gender}
\bibfield{author}{\bibinfo{person}{Sharyl~Bender Peterson} {and}
  \bibinfo{person}{Mary~Alyce Lach}.} \bibinfo{year}{1990}\natexlab{}.
\newblock \showarticletitle{Gender stereotypes in children's books: Their
  prevalence and influence on cognitive and affective development}.
\newblock \bibinfo{journal}{\emph{Gender and education}} \bibinfo{volume}{2},
  \bibinfo{number}{2} (\bibinfo{year}{1990}), \bibinfo{pages}{185--197}.
\newblock


\bibitem[\protect\citeauthoryear{Pethe, Kim, and Skiena}{Pethe
  et~al\mbox{.}}{2020}]%
        {pethe2020chapter}
\bibfield{author}{\bibinfo{person}{Charuta Pethe}, \bibinfo{person}{Allen Kim},
  {and} \bibinfo{person}{Steven Skiena}.} \bibinfo{year}{2020}\natexlab{}.
\newblock \showarticletitle{Chapter Captor: Text Segmentation in Novels}. In
  \bibinfo{booktitle}{\emph{Proceedings of the Conference on Empirical Methods
  in Natural Language Processing}}. \bibinfo{pages}{8373--8383}.
\newblock


\bibitem[\protect\citeauthoryear{Pujari, Mittal, Padhi, Jain, Jadon, and
  Kumar}{Pujari et~al\mbox{.}}{2019}]%
        {hindi}
\bibfield{author}{\bibinfo{person}{Arun~K Pujari}, \bibinfo{person}{Ansh
  Mittal}, \bibinfo{person}{Anshuman Padhi}, \bibinfo{person}{Anshul Jain},
  \bibinfo{person}{Mukesh Jadon}, {and} \bibinfo{person}{Vikas Kumar}.}
  \bibinfo{year}{2019}\natexlab{}.
\newblock \showarticletitle{Debiasing Gender biased Hindi Words with
  Word-embedding}. In \bibinfo{booktitle}{\emph{Proceedings of the 2019 2nd
  International Conference on Algorithms, Computing and Artificial
  Intelligence}}. \bibinfo{pages}{450--456}.
\newblock


\bibitem[\protect\citeauthoryear{QuillJS}{QuillJS}{2021a}]%
        {delta}
\bibfield{author}{\bibinfo{person}{QuillJS}.} \bibinfo{year}{2021}\natexlab{a}.
\newblock \bibinfo{title}{Delta}.
\newblock
\newblock
\urldef\tempurl%
\url{https://github.com/quilljs/delta/}
\showURL{%
\tempurl}
\newblock
\shownote{Accessed: 2022-02-04.}


\bibitem[\protect\citeauthoryear{QuillJS}{QuillJS}{2021b}]%
        {QuillJS}
\bibfield{author}{\bibinfo{person}{QuillJS}.} \bibinfo{year}{2021}\natexlab{b}.
\newblock \bibinfo{title}{QuillJS: a Rich Text Editor}.
\newblock
\newblock
\urldef\tempurl%
\url{https://quilljs.com}
\showURL{%
\tempurl}
\newblock
\shownote{Accessed: 2022-02-04.}


\bibitem[\protect\citeauthoryear{Ramasubramanian and Oliver}{Ramasubramanian
  and Oliver}{2007}]%
        {ramasubramanian2007activating}
\bibfield{author}{\bibinfo{person}{Srividya Ramasubramanian} {and}
  \bibinfo{person}{Mary~Beth Oliver}.} \bibinfo{year}{2007}\natexlab{}.
\newblock \showarticletitle{Activating and suppressing hostile and benevolent
  racism: Evidence for comparative media stereotyping}.
\newblock \bibinfo{journal}{\emph{Media psychology}} \bibinfo{volume}{9},
  \bibinfo{number}{3} (\bibinfo{year}{2007}), \bibinfo{pages}{623--646}.
\newblock


\bibitem[\protect\citeauthoryear{Reagan, Mitchell, Kiley, Danforth, and
  Dodds}{Reagan et~al\mbox{.}}{2016}]%
        {reagan2016emotional}
\bibfield{author}{\bibinfo{person}{Andrew~J Reagan}, \bibinfo{person}{Lewis
  Mitchell}, \bibinfo{person}{Dilan Kiley}, \bibinfo{person}{Christopher~M
  Danforth}, {and} \bibinfo{person}{Peter~Sheridan Dodds}.}
  \bibinfo{year}{2016}\natexlab{}.
\newblock \showarticletitle{The emotional arcs of stories are dominated by six
  basic shapes}.
\newblock \bibinfo{journal}{\emph{EPJ Data Science}} \bibinfo{volume}{5},
  \bibinfo{number}{1} (\bibinfo{year}{2016}), \bibinfo{pages}{1--12}.
\newblock


\bibitem[\protect\citeauthoryear{Rohrer, Sibert, and Ebert}{Rohrer
  et~al\mbox{.}}{1998}]%
        {DBLP:conf/infovis/RohrerSE98}
\bibfield{author}{\bibinfo{person}{Randall~M. Rohrer}, \bibinfo{person}{John~L.
  Sibert}, {and} \bibinfo{person}{David~S. Ebert}.}
  \bibinfo{year}{1998}\natexlab{}.
\newblock \showarticletitle{The Shape of Shakespeare: Visualizing Text using
  Implicit Surfaces}. In \bibinfo{booktitle}{\emph{Proceedings of the {IEEE}
  Symposium on Information Visualization}}. \bibinfo{publisher}{{IEEE}},
  \bibinfo{address}{Piscataway, NJ, USA}, \bibinfo{pages}{121--129}.
\newblock
\urldef\tempurl%
\url{https://doi.org/10.1109/INFVIS.1998.729568}
\showDOI{\tempurl}


\bibitem[\protect\citeauthoryear{R{\"o}siger, Schulz, and Reiter}{R{\"o}siger
  et~al\mbox{.}}{2018}]%
        {rosiger2018towards}
\bibfield{author}{\bibinfo{person}{Ina R{\"o}siger}, \bibinfo{person}{Sarah
  Schulz}, {and} \bibinfo{person}{Nils Reiter}.}
  \bibinfo{year}{2018}\natexlab{}.
\newblock \showarticletitle{Towards coreference for literary text: Analyzing
  domain-specific phenomena}. In \bibinfo{booktitle}{\emph{Proceedings of the
  Second Joint SIGHUM Workshop on Computational Linguistics for Cultural
  Heritage, Social Sciences, Humanities and Literature}}.
  \bibinfo{pages}{129--138}.
\newblock


\bibitem[\protect\citeauthoryear{Ross}{Ross}{2019}]%
        {ross2019media}
\bibfield{author}{\bibinfo{person}{Tara Ross}.}
  \bibinfo{year}{2019}\natexlab{}.
\newblock \showarticletitle{Media and stereotypes}.
\newblock \bibinfo{journal}{\emph{Journal: The Palgrave Handbook of Ethnicity}}
  (\bibinfo{year}{2019}), \bibinfo{pages}{1--17}.
\newblock


\bibitem[\protect\citeauthoryear{Scheuerman, Paul, and Brubaker}{Scheuerman
  et~al\mbox{.}}{2019}]%
        {scheuerman2019computers}
\bibfield{author}{\bibinfo{person}{Morgan~Klaus Scheuerman},
  \bibinfo{person}{Jacob~M Paul}, {and} \bibinfo{person}{Jed~R Brubaker}.}
  \bibinfo{year}{2019}\natexlab{}.
\newblock \showarticletitle{How computers see gender: An evaluation of gender
  classification in commercial facial analysis services}.
\newblock \bibinfo{journal}{\emph{Proceedings of the ACM on Human-Computer
  Interaction}} \bibinfo{volume}{3}, \bibinfo{number}{CSCW}
  (\bibinfo{year}{2019}), \bibinfo{pages}{1--33}.
\newblock


\bibitem[\protect\citeauthoryear{{\^S}ilic and Basic}{{\^S}ilic and
  Basic}{2010}]%
        {DBLP:conf/kes/SilicB10}
\bibfield{author}{\bibinfo{person}{Artur {\^S}ilic} {and}
  \bibinfo{person}{Bojana~Dalbelo Basic}.} \bibinfo{year}{2010}\natexlab{}.
\newblock \showarticletitle{Visualization of Text Streams: {A} Survey}. In
  \bibinfo{booktitle}{\emph{Proceedings of the International Conference on
  Knowledge-Based and Intelligent Information and Engineering Systems}}
  \emph{(\bibinfo{series}{Lecture Notes in Computer Science},
  Vol.~\bibinfo{volume}{6277})}, \bibfield{editor}{\bibinfo{person}{Rossitza
  Setchi}, \bibinfo{person}{Ivan Jordanov}, \bibinfo{person}{Robert~J.
  Howlett}, {and} \bibinfo{person}{Lakhmi~C. Jain}} (Eds.).
  \bibinfo{publisher}{Springer}, \bibinfo{pages}{31--43}.
\newblock
\urldef\tempurl%
\url{https://doi.org/10.1007/978-3-642-15390-7\_4}
\showDOI{\tempurl}


\bibitem[\protect\citeauthoryear{Stasko, G{\"{o}}rg, Liu, and Singhal}{Stasko
  et~al\mbox{.}}{2007}]%
        {DBLP:conf/ieeevast/StaskoGLS07}
\bibfield{author}{\bibinfo{person}{John~T. Stasko}, \bibinfo{person}{Carsten
  G{\"{o}}rg}, \bibinfo{person}{Zhicheng Liu}, {and} \bibinfo{person}{Kanupriya
  Singhal}.} \bibinfo{year}{2007}\natexlab{}.
\newblock \showarticletitle{Jigsaw: Supporting Investigative Analysis through
  Interactive Visualization}. In \bibinfo{booktitle}{\emph{Proceedings of the
  {IEEE} Symposium on Visual Analytics Science and Technology}}.
  \bibinfo{publisher}{{IEEE}}, \bibinfo{address}{Piscataway, NJ, USA},
  \bibinfo{pages}{131--138}.
\newblock
\urldef\tempurl%
\url{https://doi.org/10.1109/VAST.2007.4389006}
\showDOI{\tempurl}


\bibitem[\protect\citeauthoryear{Sterman, Huang, Liu, and Paulos}{Sterman
  et~al\mbox{.}}{2020}]%
        {sterman2020interacting}
\bibfield{author}{\bibinfo{person}{Sarah Sterman}, \bibinfo{person}{Evey
  Huang}, \bibinfo{person}{Vivian Liu}, {and} \bibinfo{person}{Eric Paulos}.}
  \bibinfo{year}{2020}\natexlab{}.
\newblock \showarticletitle{Interacting with Literary Style through
  Computational Tools}. In \bibinfo{booktitle}{\emph{Proceedings of the {ACM}
  Conference on Human Factors in Computing Systems}}.
  \bibinfo{publisher}{{ACM}}, \bibinfo{address}{New York, NY, USA},
  \bibinfo{pages}{1--12}.
\newblock


\bibitem[\protect\citeauthoryear{Tanahashi and Ma}{Tanahashi and Ma}{2012}]%
        {DBLP:journals/tvcg/TanahashiM12}
\bibfield{author}{\bibinfo{person}{Yuzuru Tanahashi} {and}
  \bibinfo{person}{Kwan{-}Liu Ma}.} \bibinfo{year}{2012}\natexlab{}.
\newblock \showarticletitle{Design Considerations for Optimizing Storyline
  Visualizations}.
\newblock \bibinfo{journal}{\emph{{{IEEE} Transactions on Visualization and
  Computer Graphics}}} \bibinfo{volume}{18}, \bibinfo{number}{12}
  (\bibinfo{year}{2012}), \bibinfo{pages}{2679--2688}.
\newblock
\urldef\tempurl%
\url{https://doi.org/10.1109/TVCG.2012.212}
\showDOI{\tempurl}


\bibitem[\protect\citeauthoryear{Tolstoy}{Tolstoy}{1998}]%
        {anna_kerreina}
\bibfield{author}{\bibinfo{person}{Leo Tolstoy}.}
  \bibinfo{year}{1998}\natexlab{}.
\newblock \bibinfo{title}{Anna Karenina}.
\newblock
\newblock
\urldef\tempurl%
\url{https://www.gutenberg.org/files/1399/1399-0.txt}
\showURL{%
\tempurl}
\newblock
\shownote{Accessed: 08/30/2021.}


\bibitem[\protect\citeauthoryear{Toshniwal, Wiseman, Ettinger, Livescu, and
  Gimpel}{Toshniwal et~al\mbox{.}}{2020}]%
        {toshniwal2020learning}
\bibfield{author}{\bibinfo{person}{Shubham Toshniwal}, \bibinfo{person}{Sam
  Wiseman}, \bibinfo{person}{Allyson Ettinger}, \bibinfo{person}{Karen
  Livescu}, {and} \bibinfo{person}{Kevin Gimpel}.}
  \bibinfo{year}{2020}\natexlab{}.
\newblock \showarticletitle{Learning to Ignore: Long Document Coreference with
  Bounded Memory Neural Networks}. In \bibinfo{booktitle}{\emph{Proceedings of
  the Conference on Empirical Methods in Natural Language Processing}}.
  \bibinfo{publisher}{Association for Computational Linguistics},
  \bibinfo{pages}{8519--8526}.
\newblock


\bibitem[\protect\citeauthoryear{Tsao}{Tsao}{2008}]%
        {tsao2008gender}
\bibfield{author}{\bibinfo{person}{Ya-Lun Tsao}.}
  \bibinfo{year}{2008}\natexlab{}.
\newblock \showarticletitle{Gender issues in young children's literature}.
\newblock \bibinfo{journal}{\emph{Reading Improvement}} \bibinfo{volume}{45},
  \bibinfo{number}{3} (\bibinfo{year}{2008}), \bibinfo{pages}{108--115}.
\newblock


\bibitem[\protect\citeauthoryear{Valentino}{Valentino}{1999}]%
        {valentino1999crime}
\bibfield{author}{\bibinfo{person}{Nicholas~A Valentino}.}
  \bibinfo{year}{1999}\natexlab{}.
\newblock \showarticletitle{Crime news and the priming of racial attitudes
  during evaluations of the president}.
\newblock \bibinfo{journal}{\emph{Public Opinion Quarterly}}
  (\bibinfo{year}{1999}), \bibinfo{pages}{293--320}.
\newblock


\bibitem[\protect\citeauthoryear{Viegas, Wattenberg, and Feinberg}{Viegas
  et~al\mbox{.}}{2009}]%
        {viegas2009participatory}
\bibfield{author}{\bibinfo{person}{Fernanda~B Viegas}, \bibinfo{person}{Martin
  Wattenberg}, {and} \bibinfo{person}{Jonathan Feinberg}.}
  \bibinfo{year}{2009}\natexlab{}.
\newblock \showarticletitle{Participatory visualization with wordle}.
\newblock \bibinfo{journal}{\emph{{{IEEE} Transactions on Visualization and
  Computer Graphics}}} \bibinfo{volume}{15}, \bibinfo{number}{6}
  (\bibinfo{year}{2009}), \bibinfo{pages}{1137--1144}.
\newblock


\bibitem[\protect\citeauthoryear{Watson, Sohn, Schriber, Gross, Mu{\~{n}}iz,
  and Kapadia}{Watson et~al\mbox{.}}{2019}]%
        {DBLP:conf/iui/WatsonSSGMK19}
\bibfield{author}{\bibinfo{person}{Katie Watson}, \bibinfo{person}{Samuel~S.
  Sohn}, \bibinfo{person}{Sasha Schriber}, \bibinfo{person}{Markus Gross},
  \bibinfo{person}{Carlos~Manuel Mu{\~{n}}iz}, {and} \bibinfo{person}{Mubbasir
  Kapadia}.} \bibinfo{year}{2019}\natexlab{}.
\newblock \showarticletitle{{StoryPrint}: an interactive visualization of
  stories}. In \bibinfo{booktitle}{\emph{Proceedings of the ACM Conference on
  Intelligent User Interfaces}}. \bibinfo{publisher}{{ACM}},
  \bibinfo{address}{New York, NY, USA}, \bibinfo{pages}{303--311}.
\newblock
\urldef\tempurl%
\url{https://doi.org/10.1145/3301275.3302302}
\showDOI{\tempurl}


\bibitem[\protect\citeauthoryear{Wise, Thomas, Pennock, Lantrip, Pottier,
  Schur, and Crow}{Wise et~al\mbox{.}}{1995}]%
        {DBLP:conf/infovis/WiseTPLPSC95}
\bibfield{author}{\bibinfo{person}{James~A. Wise}, \bibinfo{person}{James~J.
  Thomas}, \bibinfo{person}{Kelly Pennock}, \bibinfo{person}{David Lantrip},
  \bibinfo{person}{Marc Pottier}, \bibinfo{person}{Anne Schur}, {and}
  \bibinfo{person}{Vern Crow}.} \bibinfo{year}{1995}\natexlab{}.
\newblock \showarticletitle{Visualizing the non-visual: spatial analysis and
  interaction with information from text documents}. In
  \bibinfo{booktitle}{\emph{Proceedings of the {IEEE} Symposium on Information
  Visualization}}. \bibinfo{publisher}{{IEEE}}, \bibinfo{address}{Piscataway,
  NJ, USA}, \bibinfo{pages}{51--58}.
\newblock
\urldef\tempurl%
\url{https://doi.org/10.1109/INFVIS.1995.528686}
\showDOI{\tempurl}


\bibitem[\protect\citeauthoryear{Zhao, Wang, Yatskar, Ordonez, and Chang}{Zhao
  et~al\mbox{.}}{2018}]%
        {zhao2018gender}
\bibfield{author}{\bibinfo{person}{Jieyu Zhao}, \bibinfo{person}{Tianlu Wang},
  \bibinfo{person}{Mark Yatskar}, \bibinfo{person}{Vicente Ordonez}, {and}
  \bibinfo{person}{Kai-Wei Chang}.} \bibinfo{year}{2018}\natexlab{}.
\newblock \showarticletitle{Gender bias in coreference resolution: Evaluation
  and debiasing methods}.
\newblock \bibinfo{journal}{\emph{arXiv preprint arXiv:1804.06876}}
  (\bibinfo{year}{2018}).
\newblock


\bibitem[\protect\citeauthoryear{Zhou, Shi, Zhao, Huang, Chen, Cotterell, and
  Chang}{Zhou et~al\mbox{.}}{2019}]%
        {fr_es}
\bibfield{author}{\bibinfo{person}{Pei Zhou}, \bibinfo{person}{Weijia Shi},
  \bibinfo{person}{Jieyu Zhao}, \bibinfo{person}{Kuan-Hao Huang},
  \bibinfo{person}{Muhao Chen}, \bibinfo{person}{Ryan Cotterell}, {and}
  \bibinfo{person}{Kai-Wei Chang}.} \bibinfo{year}{2019}\natexlab{}.
\newblock \showarticletitle{Examining Gender Bias in Languages with Grammatical
  Gender}. In \bibinfo{booktitle}{\emph{Proceedings of the Conference on
  Empirical Methods in Natural Language Processing and the International Joint
  Conference on Natural Language Processing}}. \bibinfo{publisher}{Association
  for Computational Linguistics}, \bibinfo{address}{Florence, Italy},
  \bibinfo{pages}{5279--5287}.
\newblock


\end{thebibliography}

\end{document}